\documentclass[twocolumn,aps,prd,   
               preprintnumbers,numbers,sort&compress,
               nofootinbib,
                            showpacs,
               colorlinks,
               linkcolor=blue,
               citecolor=blue]{revtex4-1} 

\usepackage{amssymb,graphicx}
\usepackage{amsmath}
\usepackage[]{empheq}
\usepackage{hyperref}
\usepackage{natbib,ifthen}

\usepackage{wasysym}
\usepackage{mathrsfs}
\usepackage[lofdepth,lotdepth,caption=false]{subfig}
\usepackage{color}
\usepackage{bbold}

\newcommand{\exclude}[1]{}

\def\ra{\rangle}
\def\la{\langle}

\newcommand{\commentOut}[1]{}

\newcommand{\be}{\begin{equation}}
\newcommand{\ee}{\end{equation}}
\newcommand{\beq}{\begin{eqnarray}}
\newcommand{\eeq}{\end{eqnarray}}

\newcommand{\jcap}{JCAP}

\newcommand{\aap}{Astronomy \& Astrophysics}

\newcommand{\solphys}{Sol.\ Phys.}

\begin{document}
    
    
       \title{Neutron Stars as the Dark Matter  detectors  }
        \author{Ariel  Zhitnitsky}
       \email{arz@phas.ubc.ca}
       \affiliation{Department of Physics and Astronomy, University of British Columbia, Vancouver, V6T 1Z1, BC, Canada}

      \begin{abstract}
      It has been known for   quite sometime that the Neutron Stars (NS) can play a role of the Dark Matter (DM) detectors due to many uniques features of NS.  We apply these (previously developed) ideas to a specific form of the DM when it is represented by  a composite object, rather than by a local fundamental field (such as WIMPs).   To be more precise we consider the  so-called axion quark nuggets (AQN) dark matter model,   when the ``non-baryonic" dark matter  in fact  is   made of  quarks and gluons which are in dense quark phase (similar to the       old idea of the Witten's strangelets).    
   We argue that the interaction of the AQNs with NS material    may lead to many profound observable   effects, which dramatically different from conventional picture when DM particles are represented by weakly interacting   WIMPs.  In particular, we argue that the AQNs may serve as the triggers for the magnetic reconnection to heat the NS surface. This effect  may strongly  alleviate (or even completely remove) the observed inconsistencies between  the predicted and  observed surface temperatures for many old NS. This  heating mechanism is  always accompanied by the hard X ray emission, which may serve  as an indicator of the proposed mechanism.  
                    
        \end{abstract}
     
        \maketitle


	
\section{Introduction}\label{sec:introduction}
It has been known for a long time that the  dynamics of the Neutron Stars (NS)  can be   modified by the influence of the Dark Matter (DM) particles
\cite{PhysRevD.77.023006,PhysRevD.82.063531,PhysRevD.81.123521,PhysRevD.83.083512,
PhysRevD.85.023519,PhysRevD.87.123507,PhysRevD.89.015010, 
PhysRevLett.113.191301,PhysRevD.96.063002,PhysRevLett.119.131801,Bramante:2017ulk,Raj:2017wrv,Chatterjee:2022dhp}.
Furthermore, it has been also known that the basic properties of the DM particles  can be strongly constrained by considering DM-NS interactions. 
In particular, the conventional cooling pattern of NS can be modified due to the capturing and consequent  annihilation  of the DM particles \cite{PhysRevD.77.023006,PhysRevD.82.063531,PhysRevD.81.123521}. There are many other processes  which could   potentially become observable as a result of interaction of the DM particles with very dense NS environment.  We refer to several original papers \cite{PhysRevD.77.023006,PhysRevD.82.063531,PhysRevD.81.123521,PhysRevD.83.083512,
PhysRevD.85.023519,PhysRevD.87.123507,PhysRevD.89.015010, 
PhysRevLett.113.191301,PhysRevD.96.063002,PhysRevLett.119.131801,Bramante:2017ulk,Raj:2017wrv,Chatterjee:2022dhp,Bhattacharya:2023stq,Bhattacharya:2023stq} devoted to analysis of many possible physical phenomena which could   result from  such DM-NS interactions. This is obviously very broad area of research, and we refer to the  recent paper  \cite{Bramante:2023djs} for review on this topic. 

 In the present work we consider a specific model for the DM which is dramatically different from conventional paradigm when DM particles are assumed to be new (yet to be discovered)  fundamental weakly interacting massive particles (WIMPs). Before we elaborate on this specific
form of the  DM   represented by macroscopically large nuclear density  composite objects   made of   quarks and gluons  we  
detour with  overview of the most important features the DM particles  must satisfy.

Observational precision  data gathered during the last quarter  of century  
have guided the development of the so called concordance cosmological 
model $\Lambda$CDM of a flat universe, $\Omega 
\simeq 1$, wherein the visible hadronic matter represents only $\Omega_B 
\simeq 0.05$ a tiny fraction of the total energy density, see recent review \cite{Turner:2022gvw},
and interesting historical comments \cite{Bertone:2016nfn}.
  Most of the 
matter component of the universe is thought to be stored in some unknown 
kind of cold dark matter, $\Omega_{DM} \simeq 0.25$. The largest  
contribution $\Omega_{\Lambda} \simeq 0.70$ to the total density  
  is  cosmological dark energy with negative pressure, another mystery which 
  will not be discussed  in the present work.
   
 There is a fundamental difference between dark matter
  and ordinary matter (aside from the trivial difference
 dark vs.  visible). Indeed, 
 DM played a crucial role in the formation of the present  structure in the universe.  Without dark matter, the universe would have remained too  uniform to form the galaxies.  
 Ordinary matter could not produce fluctuations to create any significant  structures   because it remains tightly coupled to radiation, preventing it from clustering, until  recent epochs.   
On the other hand, dark matter, which is not coupled to photons, would permit tiny  fluctuations  
 to grow for a long, long time  before the ordinary matter decoupled from radiation.   The required material is called 
the  Cold Dark Matter, and  the obvious candidates are  the WIMPs   
  of any sort which are long-lived, cold and weakly interacting with visible hadronic material. 
The key parameter which enters all the cosmological observations is the corresponding cross section $\sigma$ to mass $M_{\rm DM}$ ratio which must be sufficiently small
to play the role of the DM as briefly mentioned above, i.e.
\be
\label{sigma/m}
\frac{\sigma}{M_{\rm DM}}\ll  1\frac{\rm cm^2}{\rm g}, 
\ee
and WIMPs obviously satisfy to the criteria (\ref{sigma/m}) to serve as DM particles. 
However, the WIMP framework which has been the dominant paradigm for the last 40 years has  failed as dozen of dedicated instruments could not find any traces of WIMPs though the sensitivity of the instruments had dramatically improved by many orders of magnitude during the last decades. 

In the present work we consider a  fundamentally  different type of the DM which is in form of the 
  dense  macroscopically large composite objects, similar to the Witten's quark nuggets  \cite{Witten:1984rs,Farhi:1984qu,DeRujula:1984axn}.   The corresponding objects   are called the axion quark nuggets (AQN) and behave as    {\it chameleons}: they  do not interact with the surrounding material in dilute environment, 
  such that the AQNs may serve as proper DM candidates as the corresponding condition (\ref{sigma/m}) is perfectly satisfied for the AQNs during  the structure formation when  the ratio $\sigma /M_{\rm AQN}\lesssim 10^{-10} {\rm cm^2}{\rm g^{-1}}$.  However, the same AQNs  become strongly interacting objects in sufficiently dense environment, such as planets and stars.  The interaction of AQNs with NS environment  also dramatically deviates from conventional WIMP -NS interactions. Therefore, many  observable consequences discussed previously \cite{PhysRevD.77.023006,PhysRevD.82.063531,PhysRevD.81.123521,PhysRevD.83.083512,
PhysRevD.85.023519,PhysRevD.87.123507,PhysRevD.89.015010, 
PhysRevLett.113.191301,PhysRevD.96.063002,PhysRevLett.119.131801,Bramante:2017ulk,Raj:2017wrv,Chatterjee:2022dhp,Bhattacharya:2023stq,Bramante:2023djs} are  dramatically modified
as a result of strong interactions of the AQNs with NS environment. 

In the present work we address a single but very important question related to the   NS cooling scenario: why is the observed surface temperature of many old NS are well above the conventional theoretical predictions? The proposed answer is that this  excess of heating is the result of the AQN-NS interaction, which is the topic of the present work. 

 Before we present our arguments supporting this claim we briefly overview several  previously proposed heating mechanisms in next Sect. \ref{NS-heating-mechanisms}  which  could in principle generate some  extra heat and could potentially answer the question formulated above.  We argue, however, that standard  astrophysical sources are very unlikely to be responsible for the observed excess of heating and there must be some unconventional sources of heat  to explain the    anomalies in observations. 
  
  The rest of our  presentation is organized as follows.  In   Sect. \ref{AQN} we overview  the basic features of the AQN model relevant for the present studies.  In Sect. \ref{corona} we  formulate the main lessons to be learned  from our previous studies of   the AQN  interactions with    solar corona.  We apply these ideas to AQN-NS interaction in Sect. \ref{NS} where we argue  that DM in form of the AQNs may serve as the triggers  igniting the  large explosive events (due to magnetic reconnections, similar to solar flares).   The dynamics of magnetic reconnection as the heating source of NS is elaborated  in details in sections \ref{NS-heating} and \ref{sect:reconnection}.   In Sect. \ref{x-ray} we argued that the study of the hard X ray emission  from NS (using Magnificent Seven stars as an example) can serve as the indicator of the proposed  heating mechanism.  We list our basic results in Sect. \ref{conclusion} where we also mention some other manifestations and observational consequences of the AQN framework. Many technical details on physics of the magnetic helicity $\cal{H}$ powering the magnetic reconnection are discussed separately in Appendix \ref{sect:helicity}.

\section{possible heating mechanisms of old NS}\label{NS-heating-mechanisms}
We start  with brief overview of the minimal cooling theory, see original paper \cite{Yakovlev:2004iq} and recent review \cite{Bramante:2023djs}.  
The observations in general are in very good agreement with minimal cooling paradigm when the neutrino emission from the core dominates at early times ($t\lesssim 10^5 \rm yr$). The  photon emission from the surface dominates  at $t\gtrsim 10^5 \rm yr$, when the neutrino emission rate gets highly suppressed  and  the NS cools down. As a result, it is expected that the NS surface temperature rapidly decreases to $T_s^{\infty}\lesssim 10^4$ K at $t\gtrsim 10^6 \rm yr$, which represents a generic consequence of the minimal cooling theory\footnote{The $T_s^{\infty}$ in this work is defined as the observed surface temperature at infinity. To be more precise, the  $T_s^{\infty}$ is defined as  $T_s^{\infty}=T_s\sqrt{1-\frac{2GM}{R}}$, where $T_s$ is the surface temperature in its local reference frame, while  $R$ and $M$ are the radius and   the mass of the NS.}. 

However, some recent studies   of old pulsars apparently are inconsistent with this canonical cooling theory, see e.g. \cite{Gonzalez:2010ta,Yanagi:2019vrr,Kopp:2022lru}   for the references on the numerous original results. These results suggest that some new sources of heating must be operational to explain the observed surface temperature being higher than expected\footnote{\label{subtleties}One should emphasize that there are numerous subtle points in ``measuring" of the NS's temperature as it is influenced by a non-thermal component. Furthermore, there are often the hot spots localized at the poles which also may dramatically modify the ``measuring" of the  average temperature of the NS. The author is thankful to the anonymous Referee for pointing out on these subtleties in  ``measuring" of the  average temperatures.}. There are many subtle points in such ``measuring"  of the surface temperatures, see footnote \ref{subtleties} with comments,  such that all recorded values  should be taken with a grain of salt and with some scepticism.

 Nevertheless, this phenomenon when ``measured" surface temperature is much higher (than naively expected temperature) is very common and generic, and it is unlikely can be entirely explained by combination of the uncertainties mentioned above and in footnote \ref{subtleties}. In fact, a higher (than expected) temperature  is observed  for very different stars with very different properties such as period, age, magnetic field, etc, which supports the claim 
that canonical cooling is far from being sufficient to explain all the observations. 
 In  principle one could  do modelling to  separate different radiation mechanisms    to extract more precisely the average value of temperature for a given NS. These questions are  well outside the scope of the present work and we refer to the   papers already mentioned  \cite{Gonzalez:2010ta,Yanagi:2019vrr,Kopp:2022lru} for review. For our arguments which follow   any precise values of the  ``measured" temperatures are not essential. Our arguments are based on a qualitative observation that in large number of  cases the observed temperature is higher than predicted, and we propose that some  accompanied effects (not directly related to measurement of the surface temperature, see Sect. \ref{x-ray}) may test our proposal.

Many mechanisms which could potentially heat the old NS have been suggested.
In particular, it includes magnetic field decay, DM accretion, crust cracking, vortex creep, roto-chemical heating, to name just a few, see   \cite{Gonzalez:2010ta} for a brief review and references. It has been argued previously that some NS which are slightly older than $\sim 10^6 \rm yr$ can be explained by some (or the combination) of these mechanisms. However, there are still many cases when suggested mechanisms are not capable to explain the data, see details below.  

In the rest of this section we critically overview   some of the most promising mechanisms suggested previously.  We also identify some cases when these  mechanisms still fail to explain the observed data. Precisely this dramatic failure in  explanation of  the observed data was the main motivation for the present work 
to suggest a novel heating mechanism which has the  potential to explain the observed anomalies. First, we start with a brief overview of previously suggested heating mechanisms, see e.g. \cite{Gonzalez:2010ta}.  

\subsubsection{Rotochemical heating}\label{sect:rotochemical}
The rotochemical heating is considered to be  the most promising heating mechanism \cite{Gonzalez:2010ta,Yanagi:2019vrr,Kopp:2022lru}. The basic idea 
is that the chemical equilibrium is altered when the rotation of the star is slowing down. The relaxing to the new equilibrium state enforces the emission  of photons which eventually heat the surface. There is a number of uncertainties   in the estimates which could be sensitive to many parameters of the models, such as the gap, the EoS,  or initial conditions for young pulsars expressed in terms of the initial period $P_0$. The fitting all these parameters in principle allows to explain the observed surface temperatures of ordinary pulsars. However, in many cases this mechanism fails to explain the observed surface temperatures.

For example, the  $T_s^{\infty}$ of the so-called Magnificent Seven stars cannot be explained by this mechanism with reasonable  changes of the parameters, see Fig. 3 in \cite{Yanagi:2019vrr}. Another set of examples includes   the old pulsars with $t\gtrsim 10^9 \rm yr$ with $T_s^{\infty}\sim 10^5$ K, see e.g. two first lines in Table 1 in   \cite{Yanagi:2019vrr}.  Such old and warm pulsars obviously cannot be   explained by  the rotochemical heating mechanism with  any reasonable modifications of the parameters, see Fig. 3 and 4 in \cite{Yanagi:2019vrr}. There are many similar cases when the observed temperatures $T_s^{\infty}$ dramatically exceeds 
the theoretical estimates, and we shall not discuss all these cases in details.

For the  purposes of the present work the most important outcome of these estimates is that the rotochemical heating is obviously could be efficient and operational in many cases. However, there are also many cases when it dramatically fails, which obviously implies that: a). there is no unified and simple mechanism which is capable to explain the observed data; b). there must be some other mechanisms which  could be also important and  which become especially pronounced at later times of the NS evolution.   
 \subsubsection{Magnetic field decay}\label{sect:magnetic}
 Another mechanism of heating which was widely discussed in the literature in the past is the heating due to the magnetic field decay, see e.g. review  \cite{Gonzalez:2010ta}.
 It is common and generally accepted view that the magnetic field cannot play a role of heating for relatively old stars with $t\gtrsim 10^6 \rm yr$. Nevertheless, we opted to present the conventional arguments (on irrelevance of the magnetic field) below  as the AQN-induced mechanism of heating of NS to be introduced later in Sect. \ref{NS} will be based precisely on the transferring the magnetic energy to the heat. We will show in Sect. \ref{sect:reconnection} where and why the conventional arguments  (on irrelevance of the magnetic field) fail for this specific AQN model. 
 
 The basic idea of the naive estimate is to assume that the decaying magnetic field strength $\cal{B}$ transfers its energy to the surface on the time scale $t$.
 In this case one can equalize the luminosity $L$ of the NS with decreasing magnetic energy in the entire NS, i.e.
 \be
 \label{B-decay}
 L=4\pi R^2\sigma T_s^4\approx \frac{4\pi R^3}{3}\cdot \frac{\la {\cal{B}}^2\ra}{8\pi}\cdot \frac{1}{t}.
 \ee
The corresponding numerical estimate \cite{Gonzalez:2010ta} suggests that the required magnetic field to explain the observations  with $T_s\sim 10^5$K  is too high. Indeed, 
\be
 \label{B-decay1}
  \sqrt{{\la {\cal{B}}^2\ra}} \sim 10^{13}\cdot \sqrt{\frac{t}{10^7 \rm yr}}\cdot\sqrt{ \frac{10 \rm~km}{R} }\cdot\left(\frac{T_s}{10^5 \rm K}\right)^2 ~~\rm G,  
 \ee
which is much higher than the observed magnetic field in classical pulsars $\sim 10^{11} G$ and millisecond pulsars $\sim 10^8$ G.
As a result of this simple estimate the   magnetic field as the source of heating was largely  ignored in the literature, in spite of the fact that the magnetic field potentially represents enormous    energy reservoir. 

There are many subtle elements in this oversimplified estimate
 because     many assumptions  being  incorporated into (\ref{B-decay}) and (\ref{B-decay1}) may not be justified. Indeed, in the estimate above it was assumed that the magnetic field is dominated by large scale dipole, which may not be the case because  very   different configurations may be the dominant contributors  to the magnetic energy. In fact, 
precisely  the enormous   magnetic energy reservoir   will play a key role in heating of the NS within the AQN framework as we argue in this work.  
The relevant configuration though is not represented in terms of  a simple large scale dipole configuration, as assumed in  (\ref{B-decay1}), but rather is represented by  complicated  
 helical fields  which will be the source of the heating as we argue below in Sect.\ref{sect:reconnection}.
Furthermore, the relevant time scale 
entering the right hand side of    estimate (\ref{B-decay}) will be very different from $(t)$ entering (\ref{B-decay})  which  dramatically modifies the  over- simplified conventional estimate (\ref{B-decay1}).  

\subsubsection{Dark Matter accretion}\label{sect:DM}
We want to mention one more heating mechanism which occurs due to the  DM accretion. This mechanism  was also widely discussed in the literature. Similar to the previous case reviewed above in Sect. \ref{sect:magnetic}  it was also thought  that the DM  cannot play any essential  role of heating for   NS with $T_s\sim 10^5$K as maximum temperature which could be   achieved by DM accretion cannot exceed $T_s\approx 3\cdot10^3$K for any reasonable parameters, see e.g. review  \cite{Gonzalez:2010ta}. Therefore, this mechanism could potentially play a role but for very old stars of age $t\gtrsim 10^8 \rm yr$. 

The basic idea of the estimate is to  observe  that the maximum possible accretion rate onto a NS is given by
\be
\label{rate}
\dot{M}\approx \rho_{\rm DM}\pi b^2_{\infty}v_{\infty}, 
\ee  
 where  $\rho_{\rm DM}$ is the local DM density in the vicinity of the NS, and   the velocity $v_{\infty}$ and  impact parameter  $b_{\infty}$ are  defined at very large distance from NS. 
 Assuming that entire energy of the DM particles is released in form of the heat one can infer that 
 \be
\label{rate1}
\dot{M}c^2\lesssim 5\cdot 10^{22}\rm \frac{erg}{s}, 
\ee 
 which is many orders of magnitude smaller than the radiation from NS with temperature $T_s\sim  10^5$K.
 Indeed, 
 \be
 \label{NS-emission}
  L=4\pi R^2\sigma T_s^4\approx  7\cdot 10^{28}\left(\frac{T_s}{10^5 \rm K}\right)^4\rm \frac{erg}{s}, 
 \ee
 which is 6 orders of magnitude higher than DM accretion heating mechanism can provide according to estimate  (\ref{rate1}). 
 As a result of this simple estimate it has been concluded  that the   Dark Matter accretion mechanism  may play a role only for very old stars  of age $t\gtrsim 10^8 \rm yr$, and  it can be safely ignored for younger NS   with temperature $T_s\gtrsim  10^5$K.
 
 The irrelevance of the DM physics for NS   with temperature $T_s\gtrsim  10^5$K was based on the canonical assumption that the DM particles are some kind of fundamental (yet to be discovered) new particles in form of WIMPs.  Precisely this type of DM particles was previously considered in the literature 
 in context of DM-NS interaction \cite{PhysRevD.77.023006,PhysRevD.82.063531,PhysRevD.81.123521,PhysRevD.83.083512,
PhysRevD.85.023519,PhysRevD.87.123507,PhysRevD.89.015010, 
PhysRevLett.113.191301,PhysRevD.96.063002,PhysRevLett.119.131801,Bramante:2017ulk,Raj:2017wrv,Chatterjee:2022dhp,Bhattacharya:2023stq,Bramante:2023djs}.

In contrast with this WIMP framework  we are advocating the AQN framework where DM is in form of the 
  dense  macroscopically large composite objects, similar to the Witten's quark nuggets  \cite{Witten:1984rs,Farhi:1984qu,DeRujula:1984axn}, as mentioned in Sect. \ref{sec:introduction}. In this case the macroscopically large AQNs can play the dual role: they obviously can inject the energy directly, similar to WIMPs,  in which case the constraint  (\ref{rate1}) holds (with minor numerical modifications). 
 
 However, the same AQNs could also play the role of the triggers which can initiate and ignite  
 magnetic reconnections such that the enormous  magnetic   energy reservoir stored in NS atmosphere and crust can heat the NS surface  
 as will be discussed in Sect.\ref{NS}.  Precisely this dual role of the AQNs dramatically modifies the conclusion of Sect. \ref{sect:magnetic} on irrelevance of the magnetic field  as a possible heating mechanism of sufficiently old NS. 
  
   Before we present the main ingredients   on the AQN-induced heating mechanism of the NS in Sects. \ref{NS}, \ref{NS-heating} and \ref{sect:reconnection} we have to make a detour to introduce the basics of the AQN framework in Sect. \ref{AQN}  and the lessons to be learned from similar processes (though in dramatically different environment) of interactions of  AQNs with solar corona in Sect \ref{corona}.

 \section{The AQN   dark matter  model }\label{AQN}
 We overview  the fundamental  ideas of the AQN model in subsection \ref{basics},  while  in subsection 
 \ref{AQN-dense} we list    some  specific features of the AQNs  relevant for the present work.
 
 \subsection{The basics}\label{basics}
 As we already mentioned 
 the AQN construction in many respects is 
similar to the Witten's quark nuggets, see  \cite{Witten:1984rs,Farhi:1984qu,DeRujula:1984axn}. This type of DM  is ``cosmologically dark'' as a result of smallness of the parameter  (\ref{sigma/m})  relevant for cosmology.  
This numerically small ratio scales down many observable consequences of an otherwise strongly-interacting DM candidate in form of the AQN nuggets.  

There are several additional elements in the AQN model in comparison with the older well-known and well-studied  {theoretical} constructions \cite{Witten:1984rs,Farhi:1984qu,DeRujula:1984axn}. First, there is an additional stabilization factor for the nuggets provided by the axion domain walls which  are copiously produced  during the   QCD  transition. This additional element    helps to alleviate a number of  problems with the original Witten's  model. In particular, a first-order phase transition is not a required feature for nugget formation as the axion domain wall (with internal QCD substructure)  plays the role of the squeezer. 

Another problem of the old construction  \cite{Witten:1984rs,Farhi:1984qu,DeRujula:1984axn} is that nuggets likely evaporate on the Hubble time-scale. For the AQN model, this is not the case because the vacuum-ground-state energies inside (the colour- superconducting phase) and outside the nugget (the hadronic phase) are drastically different. Therefore, these two systems can coexist only in the presence of an external pressure, provided by the axion domain wall, which is inevitable feature of the AQN construction. This should  be contrasted with the original model \cite{Witten:1984rs,Farhi:1984qu,DeRujula:1984axn}, which is assumed to be  stable  at zero external pressure. 
This difference has dramatic observational consequence relevant for the present work- the Witten's nugget will turn a NS  into the quark star if it hits the NS. In contrast,  a  matter type AQN   will not turn an entire star into a new quark phase because the  quark matter in the AQNs   is supported  by external axion domain wall pressure, and therefore, can be extended only to relatively small distance $\sim m_a^{-1}$,   which is much shorter  than the NS size.  

Finally,  the nuggets can be made of {\it matter} as well as {\it antimatter} during the QCD transition. 
\exclude{
The original motivation for the AQN model  can be explained as follows. 
It is commonly  assumed that the Universe 
began in a symmetric state with zero global baryonic charge 
and later (through some baryon-number-violating process, non-equilibrium dynamics, and $\cal{CP}$-violation effects, realizing the three  famous  Sakharov criteria) 
evolved into a state with a net positive baryon number.

As an 
alternative to this scenario, we advocate a model in which 
``baryogenesis'' is actually a charge-separation (rather than charge-generation) process 
in which the global baryon number of the universe remains 
zero at all times.   This  represents the key element of the AQN construction.

 In other words,  the unobserved antibaryons  in this model comprise 
dark matter being in the form of dense nuggets of antiquarks and gluons in the  colour superconducting (CS) phase.  
The result of this ``charge-separation process'' are two populations of AQN carrying positive and 
negative baryon number. The global  $\cal CP$ violating processes associated with the so-called initial misalignment angle $\theta_0$ which was present  during 
the early formation stage,  the number of nuggets and antinuggets 
  will be different.
 This difference is always an order-of-one effect irrespective of the 
parameters of the theory, the axion mass $m_a$ or the initial misalignment angle $\theta_0$.

 }
The presence of the antimatter nuggets in the AQN  framework is an inevitable and the direct consequence of the 
    $\cal{CP}$ violating  axion field  which is present in the system during the  QCD time. As a result of this feature      the DM density, 
    $\Omega_{\rm DM}$, and the visible    density, $\Omega_{\rm visible}$, will automatically assume the  same order of magnitude densities  
    $\Omega_{\rm DM}\sim \Omega_{\rm visible}$  irrespective  to the parameters of the model, such as the axion mass $m_a$. 
 This feature represents a generic property of the construction   \cite{Zhitnitsky:2002qa} as both component, the visible, and the dark are proportional to one and the same fundamental dimensional constant of the theory, the $\Lambda_{\rm QCD}$. 
  
  We refer to the original papers   \cite{Liang:2016tqc,Ge:2017ttc,Ge:2017idw,Ge:2019voa} devoted to the specific questions  related to the nugget's formation, generation of the baryon asymmetry, and  survival   pattern of the nuggets during the evolution in  early Universe with its unfriendly environment. We also refer to a recent brief review article \cite{Zhitnitsky:2021iwg} which explains a number of subtle points on the formation mechanism,  survival pattern of the AQNs during the early stages of the evolution,  including the Cosmic Microwave Background (CMB)  Big Bang Nucleosynthesis (BBN), and recombination epochs. 
  
  The only comment we would like to make here is that in this work   we take the  agnostic viewpoint, and assume that such nuggets made of {\it antimatter} are present in our Universe today   irrespective to their   formation mechanism. This assumption is consistent with all presently available cosmological, astrophysical and terrestrial  constraints as long as  the average baryon charge of the nuggets is sufficiently large as we review  below.


 \begin{table*}
\captionsetup{justification=raggedright}
	\begin{tabular}{cccrcc} 
		\hline\hline
		  Property  && \begin{tabular} {@{}c@{}}{ Typical value or feature}~~~~~\end{tabular} \\\hline
		  AQN's mass~  $[M_N]$ &&         $M_N\approx 16\,g\,(B/10^{25})$     \cite{Zhitnitsky:2021iwg}     \\
		   baryon charge constraints~   $ [B]  $   &&        $ B \geq 3\cdot 10^{24}  $     \cite{Zhitnitsky:2021iwg}    \\
		   annihilation cross section~  $[\sigma]$ &&     $\sigma\approx\kappa\pi R^2\simeq 1.5\cdot 10^{-9} {\rm cm^2} \cdot  \kappa (R/2.2\cdot 10^{-5}\rm cm)^2$  ~~~~     \\
		  density of AQNs~ $[n_{\rm AQN}]$         &&          $n_{\rm AQN} \sim 0.3\cdot 10^{-25} {\rm cm^{-3}} (10^{25}/B) $   \cite{Zhitnitsky:2021iwg} \\
		  survival pattern during BBN &&       $\Delta B/B\ll 1$  \cite{Zhitnitsky:2006vt,Flambaum:2018ohm,SinghSidhu:2020cxw,Santillan:2020lbj} \\
		  survival pattern during CMB &&           $\Delta B/B\ll 1$ \cite{Zhitnitsky:2006vt,Lawson:2018qkc,SinghSidhu:2020cxw} \\
		  survival pattern during post-recombination &&   $\Delta B/B\ll 1$ \cite{Ge:2019voa} \\\hline
	\end{tabular}
	\caption{Basic  properties of the AQNs adopted from \cite{Budker:2020mqk}. The parameter $\kappa$ in Table  is introduced to account  for possible deviation from geometric value $\pi R^2$ as a result of ionization of the AQNs due to interaction with environment. The ratio $\Delta B/B\ll 1$ in the Table implies that only a small portion $\Delta B$  of the total (anti)baryon charge $  B$  hidden in form of the AQNs get annihilated during big-bang nucleosynthesis (BBN), Cosmic Microwave Background (CMB), or post-recombination epochs (including the galaxy and star formation), while the dominant portion of the baryon charge survives until the present time.  } 
	\label{tab:basics}
\end{table*}

   \exclude{
 The strongest direct detection limit\footnote{Non-detection of etching tracks in ancient mica gives another indirect constraint on the flux of   DM nuggets with mass $M> 55$g   \cite{Jacobs:2014yca}. This constraint is based on assumption that all nuggets have the same mass, which is not the case  for the AQN model.} is  set by the IceCube Observatory's,  see Appendix A in \cite{Lawson:2019cvy}:
\be
\label{direct}
\la B \ra > 3\cdot 10^{24} ~~~[{\rm direct ~ (non)detection ~constraint]}.
\ee
The basic idea of the constraint (\ref{direct}) is that IceCube  with its surface   area $\rm \sim km^2$  has not detected any events during  its 10 years of observations. In the estimate  (\ref{direct})
it was assumed that the efficiency of  detection of a macroscopically large nugget is 100$\%$ which excludes AQNs with small baryon charges 
$\la B \ra < 3\cdot 10^{24}$ with   $\sim 3.5 \sigma$ confidence level.

Similar limits are   also obtainable 
from the   ANITA 
  and from  geothermal constraints which are also consistent with (\ref{direct}) as estimated in \cite{Gorham:2012hy}. It has been also argued in \cite{Gorham:2015rfa} that   AQNs producing a significant neutrino flux 
in the 20-50 MeV range cannot account for more than 20$\%$ of the DM 
density. However, the estimates \cite{Gorham:2015rfa} were based on assumption that the neutrino spectrum is similar to  the one which is observed in 
conventional baryon-antibaryon annihilation events, which is not the case for the AQN model when the ground state of the quark matter is in the 
colour superconducting (CS) phase, which leads to the dramatically different spectral features.  The  resulting flux computed in \cite{Lawson:2015cla} is perfectly consistent with observational constraints. 

The authors of Ref. \cite{SinghSidhu:2020cxw} considered a generic constraint for the nuggets made of antimatter (ignoring all essential  specifics of the AQN model such as quark matter  CS phase of the nugget's core). Our constraints (\ref{direct}) are consistent with their findings including the CMB and BBN, and others, except the constraints derived from    the so-called ``Human Detectors". 
As explained in \cite{Ge:2020xvf}
  the corresponding estimates of Ref. \cite{SinghSidhu:2020cxw} are   oversimplified   and do not have the same status as those derived from CMB or BBN constraints.  

 While ground based direct searches   
offer the most unambiguous channel
for the detection of the conventional DM candidates such as  Weakly Interacting Massive Particles (WIMP),  
the flux of AQNs    is inversely proportional to the nugget's mass   and 
consequently even the largest available conventional DM detectors are incapable  to exclude  (or even constrain)   the  potential mass range of the nuggets. Instead, the large area detectors which are normally designed for analyzing     the high energy cosmic rays are much better suited for our studies of the AQNs as we discuss in next section \ref{earth}. 

 }

     \begin{figure}[h]
	\centering
	\captionsetup{justification=raggedright}
	\includegraphics[width=0.8\linewidth]{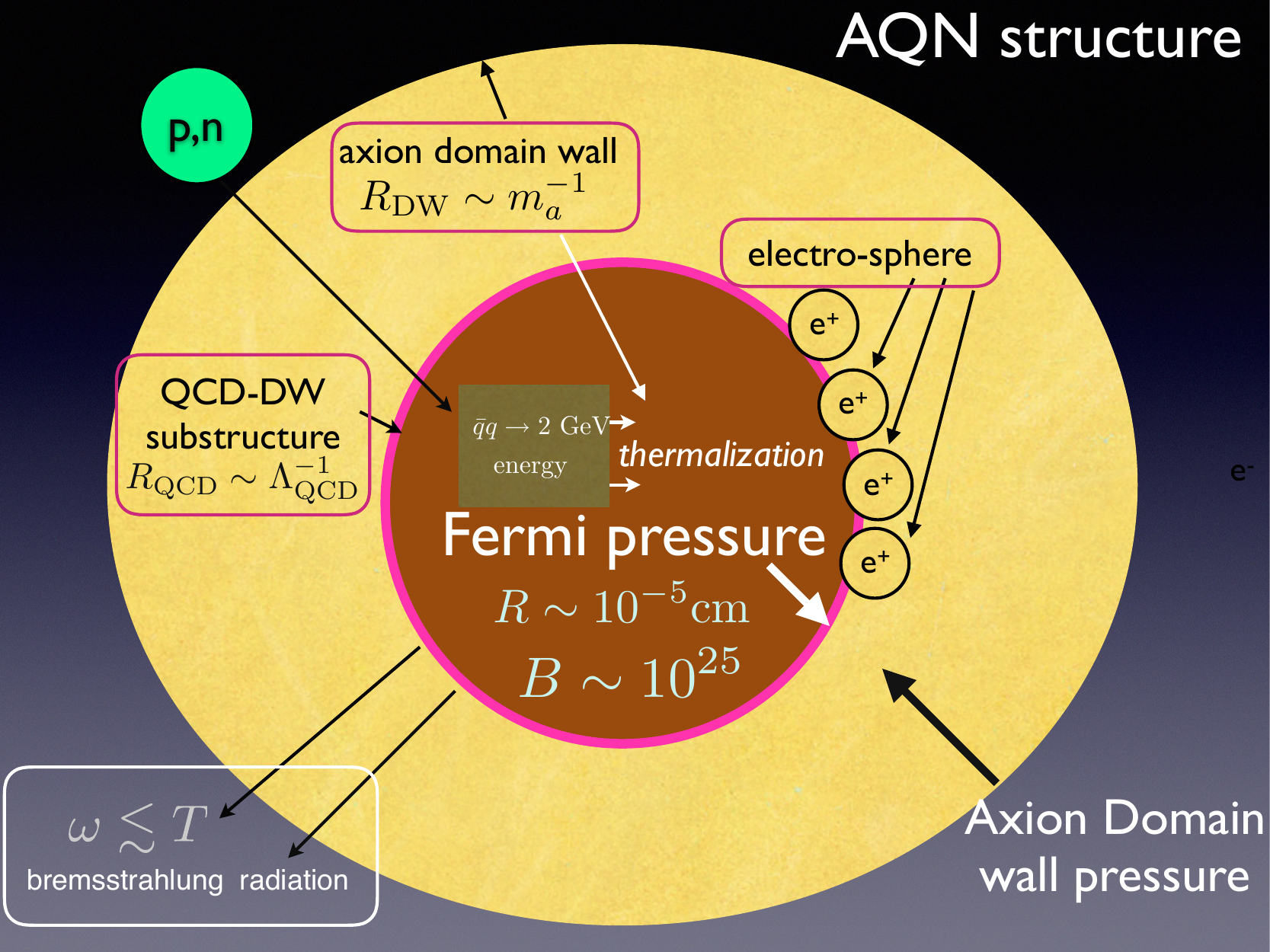}
		\caption{AQN-structure (not in scale), adopted from \cite{Zhitnitsky:2022swb}. The dominant portion of the energy $\sim 2$ GeV produced as a result of  a single  annihilation process inside the anti-nugget is released in form of the bremsstrahlung radiation with frequencies $\omega\leq T$, see description and notations in the main text.}
\label{AQN-structure}
\end{figure}

We conclude this brief review subsection with Table\,\ref{tab:basics} which summarizes the basic features and parameters of the AQNs. Important point here is that
 only a small portion $\Delta B\ll B$  of the total (anti)baryon charge $  B$  hidden in form of the AQNs get annihilated during long evolution of the Universe. The dominant portion of the baryon charge survives until the present time.
  Independent analysis \cite{Santillan:2020lbj} and  \cite{SinghSidhu:2020cxw}    also support our original claims as cited in the Table\,\ref{tab:basics} that the anti-quark nuggets survive the BBN and CMB epochs. 
 

   We draw the AQN-structure on Fig \ref{AQN-structure}, where we use typical parameters from the  Table\,\ref{tab:basics}. There are several   distinct length scales of the problem: $R\sim 10^{-5}$ cm represents the size of the nugget filled by dense quark matter with total baryon charge $B\sim 10^{25}$ in CS phase. Much larger scale  $R_{\rm DW}\sim m_a^{-1}$  describes the axion DW   surrounding the quark matter. The axion DW has the QCD substructure surrounding the quark matter and  which has typical width of order $R_{\rm QCD}\sim 10^{-13} \rm cm$. Finally, there is always electro-sphere which represents a  very generic feature of quark nuggets, including the Witten's original construction. In case of antimatter-nuggets the electro-sphere comprises the positrons.  The  typical size of the electrosphere is order of $10^{-8} \rm cm$, see below.

 \subsection{When the AQNs start to interact  with dense  environment}\label{AQN-dense}
    For our present work, however,  the  most relevant studies  are related to the effects which  may occur when the AQNs made of antimatter 
    propagate in the environment with sufficiently large visible matter density $n (r)$ such as density of the solar corona, or NS atmosphere. 
   In this case  the annihilation processes start and 
      a large amount of energy   will be injected to surrounding material, which may  be manifested in many different ways. What is more important for the present studies is that the same annihilation processes  become much more important if the AQN enters the region of highly ionized plasma because the ions are much more likely to interact  with the AQNs in comparison with neutral atoms due to the long-ranged Coulomb attraction.  
      
      The related  computations on the AQN-visible matter interaction originally have been carried out in \cite{Forbes:2008uf}
 in application to the galactic neutral environment at present time  with a typical density of surrounding   baryons of order $n_{\rm galaxy}\sim   {\rm cm^{-3}}$ in the galaxy. We review  these computations  with few additional elements which must be implemented in case of propagation of the AQN in denser and ionized environment such as NS atmosphere.

When the AQN enters the region of the baryon density $n $ the annihilation processes start and the internal temperature increases. 
  A typical internal temperature  $T$ of  the  AQNs   can be estimated from the condition that
 the radiative output   must balance the flux of energy onto the 
nugget 
\be
\label{eq:rad_balance}
    F_{\rm{tot}} (T) (4\pi R^2)
\approx \kappa\cdot  (\pi R_{\rm eff}^2) \cdot (2~ {\rm GeV})\cdot n \cdot v_{\rm AQN},  
\ee 
where $n$ represents the baryon number density of the surrounding material, and $F_{\rm{tot}}(T) $ is total  surface emissivity, see below. 
The left hand side accounts for the total energy radiation from the  AQN's surface per unit time   while  
 the right hand side  accounts for the rate of annihilation events when each successful annihilation event of a single baryon charge produces $\sim 2m_pc^2\approx 2~{\rm GeV}$ energy. If the environment is represented by neutral atoms and molecules the interaction of the AQNs with environment can be approximated by the geometrical cross section $ \pi R^2$ for macroscopically large object of size $R$. 
 However, if the surrounding material is highly ionized the effective cross section $  \pi R_{\rm eff}^2$ could be dramatically larger than the geometric value $ \pi R^2$ due to the long range Coulomb interaction as the AQN assumes a large negative charge at sufficiently high temperature $T$, see below for estimates. 
 The factor $\kappa$  in (\ref{eq:rad_balance}) accounts    for large theoretical uncertainties related to the annihilation processes 
of the (antimatter)  AQN  colliding with surrounding material.

        The total  surface emissivity due to the bremsstrahlung radiation from electrosphere at temperature   $T$ has been computed in \cite{Forbes:2008uf} and it is given by 
\begin{equation}
  \label{eq:P_t}
  F_{\rm{tot}} \approx 
  \frac{16}{3}
  \frac{T^4\alpha^{5/2}}{\pi}\sqrt[4]{\frac{T}{m}}\,,
\end{equation}
 where $\alpha\approx1/137$ is the fine structure constant, $m=511{\rm\,keV}$ is the mass of electron, and $T$ is the internal temperature of the AQN.  
 One should emphasize that the emission from the electrosphere is not thermal, and the spectrum is dramatically different from blackbody radiation.
    
    From (\ref{eq:rad_balance})  and (\ref{eq:P_t}) one can estimate a typical internal nugget's temperature  for the neutral environment when $R_{\rm eff} \approx R$  and the  density $n$ assumes a typical galactic values $n\sim \rm cm^{-3}$:   
     \be
 \label{T}
 T\sim 0.4 ~{\rm eV} \cdot \left(\frac{n }{\rm cm^{-3}}\right)^{\frac{4}{17}}\cdot\left(\frac{v_{\rm AQN}}{10^{-3}c}\right)^{\frac{4}{17}}  \cdot \kappa^{\frac{4}{17}}.
 \ee 
 \exclude{
 In case when the    surrounding material   is a highly ionized  plasma (which is relevant for our present studies of the NS atmosphere)   the parameter $R_{\rm eff}  $ effectively gets much larger  than geometrical size $R$ as the AQN, being a strongly negatively charged object, attracts more positively charged ions from surrounding material. This attraction    consequently   effectively increases the cross section  (\ref{eq:rad_balance}) and the rate of annihilation,  eventually resulting in much  higher  value of $T$.  
 }

Another  feature   which is relevant for our  present studies is the ionization  properties of the AQN itself (along with ionization of the surrounding plasma). Ionization, as usual,   occurs in a system   as a result of  the high internal temperature $T$, in which  case   a large number of weakly bound positrons  from the electrosphere   get excited and can easily leave the system. As a result, the anti-nugget assumes a negative  electric charge. Its absolute value     $Q(T) $  strongly depends on the environment (density and the temperature). 

 \exclude{
 is estimated as follows \cite{Flambaum:2018ohm}: 
\be
  \label{Q}
Q(T,R_{\rm eff}) \approx 4\pi   \int^{\infty}_{R_{\rm eff}}  n_{e^+}(r, T)  r^2 d r 
  \ee
 where $n_{e^+}(r, T) $ is the local density of positrons from electrosphere at distance $r$, which has been computed in the mean field approximation  in \cite{Forbes:2008uf,Forbes:2009wg}.

  The corresponding  effective  radius $R_{\rm eff}(T)$  which enters the expressions (\ref{eq:rad_balance}) and (\ref{Q}) can be estimated from the condition that the potential energy of the attraction  is the same order of magnitude as kinetic energy of the protons from highly ionized plasma. 
As a result  the protons from plasma can be captured by negatively charged AQN.  The condition of a successful  capturing of a proton  assumes the form: 
 \be
\label{capture1}
  \frac{\alpha Q(T,R_{\rm eff}) }{R_{\rm eff}(T)} \gtrsim \frac{m_p|\bf{v}-\bf{v}_{\rm AQN}|^2}{2}, ~~~~  \frac{m_p v^2}{2}\sim T_{\rm gas},
  \ee
where $Q(T,R_{\rm eff}) $ is estimated by  (\ref{Q}).  One should emphasize that $R_{\rm eff}(T)$ depends on both temperatures, the internal $T$ through the charge $Q(T)$ as given by (\ref{Q}) and external gas  temperature $T_{\rm gas}$ which is essentially determined by the typical velocities of particles in plasma. Important  point here is that $R_{\rm eff}(T)$ could be dramatically larger than the geometric size $R$ such that effective cross section being $\pi R_{\rm eff}^2$ could be  much  larger than geometric size $\pi R^2$ entering (\ref{eq:rad_balance}).

 As a result of this dynamical processes    the AQN   acquires a negative  electric  charge $\sim -|e|Q$  and  get   partially  ionized as a macroscopically large object of mass $M\simeq m_pB$. As we discuss below in all cases 
the ratio $eQ/M\ll 10^{-15} e/m_p$ characterizing  this object  is very tiny in comparison with proton, such that AQNs themselves do not change momentum nor trajectory under the influence of the external magnetic field. However, the charge $Q$ itself is sufficiently large  being capable to capture (with consequent possibility of annihilation) the  positively charged protons from surrounding ionized plasma.
}

Precisely this feature of ionization of the AQN   dramatically enhances the visible-DM interaction in highly ionized  sufficiently dense environment when cosmologically relevant ratio ($\sigma/M)$
from (\ref{sigma/m})  could become very large. This feature may   dramatically modify many previously obtained  results \cite{PhysRevD.77.023006,PhysRevD.82.063531,PhysRevD.81.123521,PhysRevD.83.083512,
PhysRevD.85.023519,PhysRevD.87.123507,PhysRevD.89.015010, 
PhysRevLett.113.191301,PhysRevD.96.063002,PhysRevLett.119.131801,Bramante:2017ulk,Raj:2017wrv,Chatterjee:2022dhp,Bhattacharya:2023stq,Bramante:2023djs}  which were made under assumption that   DM particles  are   fundamental    weakly interacting objects such as WIMPs. 

The emergence of very strong interaction of DM with surrounding material at the star's surface   is a direct manifestation of the AQN construction  when the dark matter  in form of the AQNs   behave as   {\it chameleon} like composite  objects.
Indeed, the AQNs are the     perfect DM  particles  in dilute environment as reviewed in Sect. \ref{basics} but become very strongly interacting objects 
in relatively dense environment.

Finally, one should mention here that the AQN model with the same set of parameters to be used in the present work may explain a number of   puzzling and mysterious observations 
 which cannot be explained by conventional astrophysical phenomena. These mysterious puzzles occur at many different scales 
in  dramatically different environments, including  BBN epoch, dark ages, as well as  galactic, Solar  and  Earth environments  at present time, see  concluding section \ref{sect:paradigm}.

 \section{Lessons from the solar corona heating puzzle}\label{corona} 
 Before we consider the dynamics of the AQN-NS interaction  in next section \ref{NS}     we would like to make a short detour in this section  to 
  overview the application of the AQN framework to the solar corona heating problem  \cite{Zhitnitsky:2017rop,Zhitnitsky:2018mav,Raza:2018gpb,Ge:2020xvf}.  The basic lesson from  the studies  \cite{Zhitnitsky:2017rop,Zhitnitsky:2018mav,Raza:2018gpb,Ge:2020xvf} is that the energy which heats the corona may come from the annihilation processes of the DM particles in form of the AQNs in solar corona.   Furthermore, the same AQNs may  play the role of {\it triggers} which may ignite the  large solar flares. In other words, the AQNs entering the solar corona may activate the magnetic reconnection    in active regions of the sun and initiate  the large solar flares. 

We shall use the corresponding  lessons from   \cite{Zhitnitsky:2017rop,Zhitnitsky:2018mav,Raza:2018gpb,Ge:2020xvf} for the case 
    when AQNs hit the NS, which is the topic of the present studies.  Precisely this analogy between the AQN dynamics in NS's atmosphere versus  solar corona will be our guiding principle in our studies of  the AQN-NS interactions in Sects. \ref{NS}, \ref{NS-heating} and \ref{sect:reconnection}.

\subsection{Solar corona heating puzzle: the observations }\label{nanoflares} 
We start with a few historical remarks. The solar corona is a very peculiar environment. Starting at an altitude of 1000 km above of the photosphere, the highly ionized iron lines show that the plasma temperature exceeds a few $10^6$ K. The total energy radiated away by the corona is of the order of $L_{\rm corona} \sim 10^{27} {\rm erg~ s^{-1}}$, which is about $10^{-6}-10^{-7}$ of the total energy radiated by the photosphere. Most of this energy is radiated at the extreme ultraviolet (EUV) and soft X-ray wavelengths. There is a very sharp transition region, located in the upper chromosphere, where the temperature suddenly jumps from a few thousand degrees to $10^6$ K. This transition layer is relatively thin, 200 km at most. This transition happens uniformly over the Sun, even in the quiet Sun, where the magnetic field is small ($\sim 1~ {\rm G}$), away from active spots and coronal holes. The reason for this uniform heating of the corona remains to be a mystery.

A possible solution to the heating problem in the quiet Sun corona was proposed in 1983 by Parker \citep{Parker}, who postulated that a continuous and uniform sequence of miniature flares, which he called ``nanoflares'', could happen in the corona. 

 The term ``nanoflare" has been used in a series of papers by Benz and coauthors
  \cite{Benz-2000,Benz-2001, Kraev-2001,Benz-2002, Benz-2003}, and many others, to advocate the idea that these small ``micro-events'' might be  responsible for the   heating of the   quiet solar corona.  In most  recent studies for the purpose of the modelling the term ``nanoflare"    describes   a generic event  for any impulsive energy release on a small  scale, without specifying its cause and its physics nature, see  review papers  \cite{Klimchuk:2005nx,Klimchuk:2017} 
 with references on  recent  activities in the field.  
 \exclude{
 In what follows, we adapt  the definition suggested in \cite{Benz-2003} and refer to nanoflares as ``micro-events" in quiet regions of the corona, to be contrasted with ``micro flares," which are significantly larger in scale and observed in active   regions.   The term ``micro-events" refers to a short enhancement of coronal emission in the energy range of about $(10^{24}-10^{28})$erg. One should emphasize that the lower limit   gives the instrumental threshold for observing quiet  regions, while the upper limit refers to the smallest events observable in active regions.  In other words,  in most studies  
     the hydrodynamic consequences of impulsive heating (due to the nanoflares)  have been used without discussing their     nature. In contrast, in our  work we do not discuss the hydrodynamics, but rather we address precisely the question on the origin of the energy  responsible for coronal heating problem.
 Our total  energy budget estimate (\ref{total_power})   in the AQN model suggests that  the nuggets  might be responsible for 
 coronal heating  as the corresponding estimate  is  consistent with the observed luminosity (\ref{estimate}).
 }
 The list below shows the most important constraints on 
nanoflares from the observations of the EUV iron lines with SoHO/EIT:\\
 {\bf 1.} The EUV emission is highly isotropic \citep{Benz-2001, Benz-2002}, therefore the nanoflares have to be distributed very ``uniformly in quiet regions'',  in contrast with flares which have a highly non-isotropic spatial distribution because they are associated with small  active regions;\\
  {\bf 2.} According to \cite{Kraev-2001}, in order to reproduce the measured EUV excess, the observed range of nanoflares  needs to be extrapolated from the observed events  interpolating between $(3.1\cdot 10^{24}  - 1.3\cdot 10^{26})~{\rm erg}$ to sub-resolution events with much smaller energies, see item 3 below.\\
 {\bf 3.} In order to reproduce the measured  radiation loss, the observed range of nano flares   needs to be extrapolated to  energies as low as  $ 10^{22}$erg  and in some models, even to $ 10^{20}$erg (see table 1 in \cite{Kraev-2001});\\
{\bf 4.} The  flares, in contrast with nano-flares,  originate at sunspot areas, with locally large magnetic fields ${\cal{B}}\sim (10^2-10^3)$ G, while  the EUV emission (which is observed  even in very quiet regions where ${\cal{B}}\sim 1$G) is isotropic and covers the entire solar surface;\\
\exclude{
{\bf 5.} Time measurements of many nanoflares demonstrate a Doppler shift with typical velocities of (250-310) km/s (see Fig.5 in \cite{Benz-2000}).
The observed line width in OV  of $\pm 140$ km/s far exceeds the thermal ion velocity, which is around 11 km/s   \cite{Benz-2000};\\
}
 {\bf 5.} The temporal evolution of flares and nanoflares also appears different. The typical ratio between the maximum and minimum EUV irradiance during  the solar cycle does not exceed a factor of 3 between its maximum in 2000 and its minimum in 2009 (see Fig. 1 from \cite{Bertolucci-2017}), while the same ratio for flares and sunspots is much larger, of the order of $10^2$.
If the magnetic reconnection (as Parker originally conjectured) was fully responsible for both the flares and nanoflares, then  the variation during the solar cycles should be similar for these two phenomena. It is not what is observed: the modest variation of the EUV with the solar cycles in comparison to the flare fluctuations suggests  that the EUV radiation does not directly follow the magnetic field activity, and that the EUV fluctuation is a  secondary, not a primary effect of the magnetic activity.
\exclude{
 The nanoflares are usually characterized by the following distribution:
 \begin{equation}
\label{dN}
 d N\propto   W^{-\alpha} d W,  ~~~~ 10^{21}{\rm erg}
 \lesssim W\lesssim 10^{26} {\rm erg},
\end{equation}
 where $ d N$ is the number of nanoflare events   with an energy between $W$ and $W+ d W$. 
 In formula (\ref{dN}), we display  the approximate energy window for $W$ as expressed by items {\bf 2} and {\bf 3}, including the sub-resolution events extrapolated to very low energies. 
 The distribution   $  d N /d W$ has been  modelled via magnetic-hydro-dynamics (MHD) simulations \cite{Pauluhn:2006ut,Bingert:2013} in such a way that the Solar observations match the simulations. The parameter $\alpha $ was  fixed to fit  observations \cite{Pauluhn:2006ut,Bingert:2013}.Many different models have been suggested to describe the data, including different slop parameter $\alpha$ for different segments of the entire window (\ref{dN}). However,   for our brief overview in this section  it sufficient to  use  a   very simplified  model with a single  $\alpha\approx 2$ for  the entire energy range (\ref{dN}).
 }
 
 \subsection{The nanoflares as the AQN annihilation events}\label{identification} 
 All the puzzles  (such as isotropic features  of the EUV emission over the entire solar surface  with very modest variations during the solar cycles)  as mentioned  above can be  naturally understood if the EUV emission from the solar corona is related to the DM particles. 
 The corresponding conjecture  
 that the nanoflares heating the corona can be identified with AQN annihilation events\footnote{\label{time-scales}In this sense our  proposal fulfills a key missing ingredient on the nature of the  nanoflares as conventional scaling arguments suggest that the typical time scales for the magnetic reconnection in the background of  a typical magnetic field   ($\sim 1$ G), must be very long, dramatically longer than  the observations suggest, see   Appendix in \cite{vanWaerbeke:2018nyj} for the details.}  has been explicitly formulated  in  \cite{Zhitnitsky:2017rop}. 
 The main argument supporting this conjecture  is amazing numerical coincidence  between the observed total luminosity ($\sim 10^{27} {\rm erg\cdot s^{-1}}$) radiated from corona in form of the soft X rays and EUV and the 
   injected energy  resulting from   the   annihilation events when the AQNs hit the Sun. 
 
 Indeed, the impact parameter for capture of the nuggets by the Sun can be estimated as
  \be
  \label{capture}
  b_{\rm cap}^{\odot}\simeq R_{\odot}\sqrt{1+\gamma_{\odot}}, ~~~~ \gamma_{\odot}\equiv \frac{2GM_{\odot}}{R_{\odot}v^2},
  \ee
  where $v\simeq 10^{-3}c$ is a typical velocity of the nuggets.    
    Assuming that $\rho_{\rm DM} \sim 0.3~ {\rm GeV cm^{-3}}$ and using the capture impact parameter (\ref{capture}), one can estimate 
  the total energy being injected due to the complete annihilation of the nuggets in solar corona as follows:
   \be
  \label{total_power}
   L_{\odot }^{\rm AQN}\sim (\pi b^2_{\rm cap})\cdot  v \cdot\rho_{\rm DM}   
  \simeq  10^{30}  \frac{\rm GeV}{\rm  s}\simeq  10^{27} \frac{\rm erg}{\rm  s}, 
  \ee
   where we substitute  constant $v\simeq 10^{-3}c$  for numerical  estimate\footnote{\label{fraction}A more proper  estimation should include $n_{\rm AQN}(2 ~m_p\cdot B\cdot  {3}/{5})$ where coefficient $3/5$ reflects the portion of the anti-nuggets,  $n_{\rm AQN}\approx \rho_{\rm DM}/M_N$ is the density of the AQNs while $M_N\approx m_pB$ is their typical mass. We ignore all these numerical coefficients in estimate (\ref{total_power}) in front of $\rho_{\rm DM} $.}.  Precisely this ``accidental  numerical coincidence" between the observed luminosity $L_{\rm corona} \sim 10^{27} {\rm erg~ s^{-1}}$ and the AQN-induced luminosity  (\ref{total_power}) was the main motivation   to put forward the idea that  the AQNs represent a new source of energy feeding the EUV radiation from solar corona,   which is very hard to explain in terms of conventional astrophysical sources as highlighted above in section \ref{nanoflares}, see also footnote \ref{time-scales} for a comment. 
 
 Based on this amazing numerical coincidence the nanoflares have been identified with the AQN annihilation events (within the AQN framework).
 \exclude{
 From this identification, it follows that the  baryon charge distribution   and the nanoflare energy distribution (\ref{dN}) must be one and the same function \cite{Zhitnitsky:2017rop}, i.e.
  \begin{equation}
\label{W_B}
 d N\propto B^{-\alpha} d B\propto W^{-\alpha} d W
\end{equation}
where $d N$ is the number of nanoflare events  (\ref{dN}), which occur as a result of the complete annihilation of the antimatter AQN carrying a baryon charges between $B$ and $B+ d B$.  Important point here is that power index $\alpha$ is identically the same for (naively) different distributions. This represents the formal realization of the identification  (between nanoflares and AQN annihilation events) as formulated above. 
}
An immediate self-consistency check of this conjecture is   that the lower limit    for the AQN baryonic charge (see Table \ref{tab:basics})   approximately coincides with nanoflare's low energy. Indeed, according to this identification  the  AQN annihilation  of baryon charge $B$  produces  the  energy  $W\simeq 2m_pc^2  B$. One can check that the smallest AQN baryonic charge $B\sim 10^{24}$ as given in  Table \ref{tab:basics} is indeed   close to the lowest   nanoflare's   energy $W\sim 10^{21} \rm erg$.
 We emphasize that this numerical similarity   represents a  highly nontrivial self-consistency check of proposal  \cite{Zhitnitsky:2017rop}, as the acceptable  range  for the AQNs and nanoflares have been constrained from dramatically  different  physical systems. 
  
  Encouraged by this self-consistency check and the highly nontrivial energetic consideration,  a full scale of  the Monte Carlo (MC) simulations had been performed in \cite{Raza:2018gpb}. It has been shown that the annihilation mostly occurs 
  at the altitudes around 2000 km where the most of the injected energy is released. This represents a highly nontrivial explanation of the emergence  of a very narrow transition region of order 200 km  width   within the AQN framework.

To summarize the proposal on identification of the AQN annihilation events with nanoflares:  the uniformity of the EUV emission is  naturally understood  in the AQN framework as DM is   distributed very uniformly over the Sun, making no distinction between quiet and active regions. Furthermore, our proposal explains   an  insignificant role   of the magnetic field for the   EUV radiation  as   the AQN events  do not depend on the strength of the magnetic field, which is also consistent with observations. It should be contrasted with original  conjecture \citep{Parker} where nanoflares are thought to be scaled down configurations of their larger cousins, which are known to be localized exclusively in the area with large  magnetic field, and fed  by the magnetic field energy. 
 
   Finally,  the temporal modulation of the EUV irradiance over a solar cycle is very modest, as opposed to the very dramatic changes  in flare activity  on the level of $10^2$ over the same time scale.   This is perfectly consistent with our interpretation of nanoflares being associated with AQN annihilation events which are not related to the solar activity, nor to dynamics of the magnetic field itself during the cycles. 

\subsection{AQNs as the  triggers of large solar flares}\label{AQN-flares}

In this section we  overview the basic results from \cite{Zhitnitsky:2018mav} where it was  argued that the same AQNs (which are identified with nanoflares as overviewed in previous section \ref{identification}) could serve as the   {\it triggers} for large Solar flares. 
The basic reason for necessity for  a trigger  (which can initiate the magnetic reconnection) to be present in the system    is related to very    large numerical values of   the so-called the Lundquist number
  $S\sim (10^{12}-10^{14})$  in solar corona. Precisely this parameter   $S$  determines the theoretical value for reconnection time which is much longer  than observations show, see Appendix in \cite{vanWaerbeke:2018nyj}  for  details and references.  Though in the last 10-15 years  many  new ideas have been pushed forward to speed up the reconnection, a large number of fundamental questions remains\footnote{\label{times}One of the question is the observation of the dramatic variation of time scales  when  pre-flare X ray radiation lasts for seconds,   flare itself lasts for about an hour while the preparation phase of the magnetic configurations to be reconnected could last for months. The presence of a trigger may automatically resolve this and many other hard questions, see  \cite{Zhitnitsky:2018mav}.}.

The basic idea of \cite{Zhitnitsky:2018mav}  is as follows:
\exclude{The AQNs entering the Solar Corona may activate the magnetic reconnection of {\it preexisting}  magnetic configurations in    active regions (distributed very non-uniformly).  Technically the effect  occurs  due to the {\it shock waves} which form as the typical velocity of the   nuggets $v\sim (600-800)~ {\rm km/s}$ in the vicinity of the surface, is well above the speed of sound, $v/c_s\gg 1$. The energy of the flares in this case is determined  by the  preexisted magnetic field configurations occupying very large area in active region, while the relatively small amount of energy associated with initial AQNs plays a minor role in the total  energy released during a large flare.
}

1. The AQNs entering the solar corona with  typical velocity of the   nuggets $v\sim (600-800)~ {\rm km/s}$ in the vicinity of the surface will inevitably  generate shock waves as the typical  velocities $v$ of the dark matter particles  much larger than  the  speed of sound $c_s$  such that the  Mach number $M\equiv v  /c_s >1$, see estimates below;

 2. When the AQNs (distributed uniformly) enter regions with a strong magnetic field  in {\it active regions}, they trigger magnetic reconnection of {\it preexisted} magnetic configurations;

 3. Technically, the AQNs are capable sparking magnetic reconnections due to the large discontinuities of the pressure $\Delta p/p\sim M^2$ and temperature $\Delta T/T\sim M^2$ when the AQN-induced shock front passes through the magnetic reconnection regions, see estimates for these parameters below.

\exclude{
 1. AQNs entering the solar corona will inevitably  generate shock waves as  the  speed of sound $c_s$ is smaller than
 the typical  velocities $v$ of the dark matter particles (Mach number $M=v/c_s> 1$);

 2. When the AQNs (distributed uniformly) enter regions with a strong magnetic field, they trigger magnetic reconnection of {\it preexisted} magnetic fluxes in {\it active regions};

 3. Technically, the AQNs are capable sparking magnetic reconnections due to the large discontinuities of the pressure $\Delta p/p\sim M^2$ and temperature $\Delta T/T\sim M^2$ when the shock front passes through the magnetic reconnection regions;

 4. The energy of the flares $E_{\rm flare}\sim L^2_{\perp}$ is powered  by the   magnetic field occupying very large area  $\sim L^2_{\perp}$ in active regions;

 5. The above mechanism, with few additional assumptions, predicts that the number of flares $dN(E_{\rm flare})$ with energy of order $E_{\rm flare}$ scales as $dN\sim E_{\rm flare}^{-1}$, consistent  with observations;

 6. This proposal   naturally resolves the problem of drastic separation of scales when a class-A pre-flare (measured by X-rays) lasts  for seconds, a flare itself lasts about an hour,  while the preparation phase of the magnetic configurations (to be reconnected during the flare) may last months. These three  drastically different time-scales can peacefully coexist in our framework because the trigger is not an internal part of the magnetic reconnection's dynamics.


}
Now we  estimate the relevant parameters  suggesting  that the AQNs  indeed could serve as the    triggers   igniting   and initiating the  large solar flares. 

 We start our     estimate with  the speed of sound $c_s$ in the corona at $T\simeq 10^6 K$,
 \beq
 \label{sound}
 \left(\frac{c_s}{c}\right)^2 &\simeq  &\frac{3 T\cdot\Gamma }{m_p c^2},  ~~~ c_s\simeq 7\cdot 10^{-4} c\cdot \sqrt{ \frac{T}{10^6~ {\rm K}}} 
 \eeq
 where $\Gamma=5/3$ is a specific heat ratio, $c$ is the speed of light,  and we approximate the mass density $\rho_p$ of plasma by the proton's number density density $n$ as follows $\rho_p\simeq n m_p$. The crucial observation here is that the Mach number $M$ is always  much  larger than one for a typical dark matter velocities at the surface: 
 \be
 \label{Mach}
 M\equiv \frac{v }{c_s}\simeq 4  \sqrt{ \frac{10^6~ {\rm K}}{T}} > 1.
 \ee
 As a result, a strong shock waves will be generated when the AQNs enter the solar corona.
  In the limit when the thickness of the shock wave can be ignored,  the corresponding formulae for  the discontinuities of the  pressure $p$, temperature $T$, and the density $\rho_p$ are well known and given by, see e.g. \cite{Landau}:
  \beq
 \label{shock1}
 \frac{{\rho_p}_2}{{\rho_p}_1} &\simeq  &\frac{(\Gamma+1) }{(\Gamma-1)}, ~~~~~  \frac{p_2}{p_1}\simeq M^2\cdot \frac{2\Gamma }{(\Gamma+1)} \nonumber \\
 \frac{T_2}{T_1} &\simeq  &M^2\cdot \frac{2\Gamma(\Gamma-1) }{(\Gamma+1)^2}, ~~~~ M \gg 1,
 \eeq
 where we assume $M\gg 1$ and keep the leading terms only in the corresponding formulae.

Another  important parameter which determines   importance of the magnetic pressure in comparison with kinetic pressure is  dimensionless  parameter $\beta$: 
 \beq
 \label{beta}
 \beta\equiv \frac{8\pi  p}{{\cal{B}}^2} \sim 0.05  \left(\frac{n}{10^{10} ~{\rm cm^{-3}}}\right) \left(\frac{T}{10^6 K}\right) \left(\frac{100~ G}{{\cal{B}}}\right)^2~~~~~
 \eeq
 where for numerical estimates we use typical parameters for the active regions in corona when $\beta\ll 1$.  The same relation (\ref{beta}) also explicitly shows that the magnetic field cannot play any essential role (including a very unlikely   possibility of  magnetic reconnection) outside the active regions when typical value of ${\cal{B}}\sim 1 ~G$ and $\beta\gg 1$, which was the topic of the EUV radiation in previous sections \ref{nanoflares} and \ref{identification}.

\exclude{
In the active regions   where magnetic field plays a key role, another parameter of the system is  the Alfv\'{e}n velocity  $v_A$   which assumes the following typical numerical values in the  corona environment
 \beq
 \label{alfven}
 \frac{v_A}{c}&=&\frac{{\cal{B}}}{c\sqrt{4\pi\rho_p}} \sim  2\cdot 10^{-3} \left(\frac{{\cal{B}}}{100~ G}\right)\left({ \frac{10^{10} ~{\rm cm^{-3}}}{n} }  \right)^{\frac{1}{2}}.~~~~~
  \eeq
For these parameters the numerical value for  $v_A$ is approximately three times larger than   the speed of sound $c_s$ according to (\ref{sound}).  

 One can introduce the so-called Alfv\'{e}n Mach number $M_A$ which plays a role similar to the Mach number and  defined as follows
 \be
 \label{M_A}
 M_A\equiv \frac{v}{v_A}.
 \ee
 Some finite fraction of the nuggets will have $M_A > 1$ along with large Mach number $M> 1$.   In this case one should expect that  the fast shock   develops which may trigger the large flare.   The key elements suggesting that the fast shock  wave may drastically modify  the rate of magnetic reconnection
 is based on observation that the pressure,  the temperature and the magnetic pressure  may   experience some dramatic changes (\ref{shock1}) when the shock wave approaches the reconnection region.  
}
The main  assumption for the  estimates presented above is that the AQNs can be treated as the  macroscopically large objects such that conventional classical hydrodynamics applies. In different words,  the size of the objects must be much larger than the average distance $a$ between the  particles  in surrounding plasma, which itself is determined by the density $a\approx n^{-1/3}$. This condition is perfectly satisfied for the AQNs because their  effective size $R_{\rm eff}\gg 1 \rm ~cm$ is indeed much larger than the geometric size of the quark nugget $R\sim 10^{-5} \rm cm$ due to very strong interaction with surrounding ionized plasma when positively charged ions are captured by the negatively charged nugget, see   \cite{Raza:2018gpb} for the estimates of parameter $R_{\rm eff}$ which indeed satisfies $R_{\rm eff}\gg a$ for typical density in corona.

The idea   that the shock waves may dramatically  increase the rate of magnetic reconnection is not new, and has been discussed previously in the literature,   though a quite different context, see references in Appendix  in \cite{vanWaerbeke:2018nyj}.
The new element which was  advocated in
    \cite{Zhitnitsky:2018mav}
   is that the small shock waves  resulting from  entering AQNs are widespread and generic events in the solar corona.  
 These small events identified with nanoflares  and the corresponding annihilation energy is sufficient to heat  corona as estimated by (\ref{total_power}).  However, they  do not generate large flares as $\beta\gg1$ 
   in quiet regions.
   
    The situation becomes very different however 
   if the nuggets hit the active regions with  strong magnetic field ${\cal{B}}\sim 10^2 ~G$ and $\beta\ll 1$. In this case   the AQN- induced shock waves  may ignite large flare.  This proposal may answer many questions of the complicated dynamics of the flares, including a dramatic variation of the  time scales as mentioned in footnote \ref{times}. 
  
\subsection{Energetics of   large solar flares} \label{m_reconnection}

Now we want to estimate the size and the energy scales associated with  such events. We consider separately two different stages.  First, we estimate  the scales related to the initial phase of the evolution when the AQNs produce the shock waves,  but the magnetic reconnection has not started yet.  The estimation for the second phase assumes that the magnetic reconnection, leading to a large solar flare,   is already  fully developed.  
 
 In the first, initial stage of the evolution, the magnetic reconnection has not started yet, and entire energy is related to the shock wave, which itself forms as a result of AQN entering the solar atmosphere from outer space. In this case a typical time scale when AQN completely annihilates its baryon charge is   of order of $\tau\sim 10~ {\rm sec}$, see \cite{Zhitnitsky:2017rop,Raza:2018gpb}. A typical length scale is determined by the initial velocity   of the AQN which is of order $v_{\rm AQN}\sim  (600 -700){\rm km/s}$ such that $L\sim    v_{\rm AQN}\cdot \tau \sim 5\cdot 10^3~ {\rm km}$. At the same time, a typical radius $R$ of the cone formed by the shock wave is determined by the speed of sound $c_s$, such that $R\sim M^{-1} L$, where Mach number $M$ is estimated in (\ref{Mach}). For numerical estimates below we take $M\simeq 10$. The affected  area $A$ due to the shock wave (where the magnetic reconnection starts) is estimated as $A\sim M^{-1}L^2$. 
 We  summarize the parameters of the initial stage as follows
 \beq
 \label{initial}
 \tau&\approx& 10 ~ {\rm sec}, ~~ L\approx     v_{\rm AQN}\cdot \tau \approx  5\cdot 10^3~ {\rm km} ~~ \nonumber \\
 R &\approx&   M^{-1} L\approx10^{-1}L  , ~~ A\approx R\cdot L\sim 10^{-1}L^2.
 \eeq
 We are now in position to estimate the typical energetic characteristics of the system during this {\it initial} stage. 
 The key element is the observation that the temperature $T$ experiences a large discontinuity resulting form the shock according to (\ref{shock1}). Therefore, we estimate a typical internal temperature  of the nugget $T_2$  as follows
 \be
 \label{initial1}
 \frac{T_2}{T_1}\approx  M^2\approx  10^2,  ~~~ T_2\approx M^2 T_1\approx  10^8~ {\rm K}
 \ee
  where $T_1\sim 10^6 K$ corresponds to unperturbed temperature of the solar corona before the shock passage through the area. 
  \exclude{
  One can estimate the   energy disturbance related to the shock wave $ E_{\rm shock}$ during the initial stage  of evolution 
  as follows  
  \be
 \label{initial2}
 E_{\rm shock}&\sim& n\Delta T V\sim 10^{28}{\rm erg}\cdot
  \left(\frac{M  }{10}\right)^2\cdot \left(\frac{n}{10^{10} ~{\rm cm^{-3}}}\right)\cdot\left(\frac{T_1}{10^6 K}\right)\cdot \left(\frac{V_{\rm shock}}{10^9{\rm km^3}}\right), 
 \ee
One should emphasize that $ E_{\rm shock}$ should not be interpreted as  the extra energy   generated by the AQN.
In fact,  $E_{\rm shock}$ is many orders of magnitude larger than the energy associated with AQN itself, which carries a  typical  nanoflare energy reviewed in section \ref{nanoflares}. 
 Instead, $E_{\rm shock}$  should be interpreted as a result of the redistribution of the temperature, pressure and the density between different regions inside the volume $V_{\rm shock}$ when there are regions with higher than unperturbed pressure $p_1$ and temperature $T_1$, and there are  regions with the lower than unperturbed values, while truly total energy in the volume $V_{\rm shock}$ remains almost the same,  before the shock passage. 
 }
 
 Important comment here is that  
  formula (\ref{initial1})   shows    that  the AQN's internal temperature  could reach  very high values on the level
  $ T\sim 10^8~ {\rm K}$. As a result, the AQNs could be 
 the source of    the $(1-10)~ {\rm keV}$  x-rays. Interestingly enough, the x rays emission   had  indeed  been  recorded for many large flares  few moments (pre-cursor)  before 
 the   flare starts \cite{Bruevich_2017}. Furthermore, one can explicitly see that pre-flare enhancement propagates from higher levels of corona into lower   corona and chromosphere \cite{Bruevich_2017}.   It is very hard to explain  such X rays emission pattern within conventional MHD.  In our framework the X ray emission before the large flare starts is  a natural  consequence of the proposal when the AQNs (moving from outer space  to the surface) generate the shock and play the role of the triggers initiating and igniting    large flares.
 We further comment on similarity of the x ray emission from solar corona during the flare and from NS in Sect. \ref{x-ray}.
   
   The {\it second stage} of the flare (after the {\it initial stage} described above ends)  in this framework is represented by  the magnetic reconnection   ignited by the shock wave.   We have nothing new to say about this conventional  phase of the evolution. We present the corresponding estimate  below  for the total flare's energy for completeness and following discussions, 
 \beq
 \label{developed}
  W_{\rm flare}\sim \frac{{\cal{B}}^2V_{\rm flare}}{8\pi} \sim  3 \cdot10^{30}    (\frac{{\cal{B}}}{300 \rm G})^2  (\frac{V_{\rm flare}}{10^{13} {\rm km^3}})~ {\rm erg},~~
 \eeq
 where $  V_{\rm flare}\approx L_{\perp}^2L$ with $L\sim 5\cdot 10^3 ~{\rm km}$ being   a typical length scale (\ref{initial}) where shock waves develops in solar corona\footnote{to simplify  the estimates we assume the nugget trajectory is perpendicular to the solar surface such that $L$ is oriented along  $z$ direction.}, while  $L_{\perp}^2$ is the area within  active region (sunspots) which eventually becomes a part of magnetic reconnection producing the large flares. Numerically $L_{\perp}\sim (10^3-10^4)~{\rm km}$ for microflares, and it could be as large as $L_{\perp}\sim 10^5~{\rm km}$ for large flares.  It is assumed that precisely this region of volume $V_{\rm flare}=L_{\perp}^2L$ with large average magnetic field ${\cal{B}}\sim {300~ \rm G}$ feeds the solar flare as a result of magnetic reconnection. It is also known that the magnetic reconnection is always accompanied by presence of non-trivial topological structures which   manifest   of a variety of  complex processes during the flare, see \cite{Zhitnitsky:2018mav} for references.
 
  It is quite obvious that the energy (\ref{developed}) of a fully developed flare is many orders of magnitude larger  than  the initial energy of the AQN which serves as  a trigger of a large flare. Nevertheless, this  initial  stage in  the flare evolution plays  a key role in future development   of the system  because it provides   a  very strong impulse with $\Delta T/T\gg1$ and $\Delta p/p\gg1 $ in very small and very localized area for very short period of time (\ref{initial}) in the region where the  magnetic reconnection eventually develops. Precisely the presence of a trigger explains a large number of puzzles related to dramatically different time scales which are known to exist in the system, see \cite{Zhitnitsky:2018mav} for references.

  \exclude{Indeed, this proposal   naturally resolves the problem of drastic separation of scales when a class-A pre-flare (measured by X-rays) lasts  for seconds, a flare itself lasts about an hour,  while the preparation phase of the magnetic configurations (to be reconnected during the flare) may last months. These three  drastically different time-scales can peacefully coexist in our framework because the trigger is not an internal part of the magnetic reconnection's dynamics. This proposal  also gives very reasonable estimate for frequency of appearance of the solar flares as a function of energy, see  \cite{Zhitnitsky:2018mav} for the details. 
 }

We conclude this section on solar corona heating puzzle with the following remark. Our main topic of this work is  analysis of possible effects which may occur 
when the AQNs   hit   NS surface. We shall use many  lessons   from the present  section  as the magnetic reconnection in NS triggered and initiated by AQNs may generate  many profound effects as we will discuss in next section \ref{NS}.

\section{When AQN hits the NS}\label{NS}

\subsection{Energy injection due to the AQN's annihilation  }\label{sect:AQN-energy}
First of all, we would like to estimate a total power being injected as a result of AQN hitting the NS surface and get annihilated  along its path  close to the surface.
For simplicity we ignore  the DM velocity distribution and assume that   $v_{\infty}\simeq 10^{-3}c$ is a typical velocity of the nuggets at a large distance from NS. In this case 
 the impact  parameter for  capture and  consequent annihilation  of the AQNs by  NS can be estimated as follows,
  \beq
  \label{capture1}
  \frac{b_{\rm cap}^{\rm NS}}{ R_{\rm NS}} 
\simeq \frac{c}{v_{\infty}}\cdot \left(\frac{\rm 10^6~cm}{R_{\rm NS}}\right)^{1/2} \cdot \left(\frac{M_{\rm NS}}{2~M_{\odot}}\right)^{1/2} 
  \eeq
  which replaces formula (\ref{capture}) estimated for the Sun.  The total energy being injected due to the complete annihilation of the nuggets in NS as follows:
   \be
  \label{total_power_NS}
   L_{\rm NS}^{\rm AQN}\sim (\pi b^2_{\rm cap})\cdot  v_{\infty} \cdot\rho_{\rm DM}   
  \simeq    10^{23}  \frac{\rm erg}{\rm  s}, 
  \ee
which replaces formula (\ref{total_power}) with estimated for the Sun, see also footnote \ref{fraction} with a comment. The estimate (\ref{total_power_NS}) of course gives the same order of magnitude for the AQN model as for any other DM model, see e.g.  \cite{PhysRevLett.119.131801}  where only kinetic energy of a DM particle   contributes to the heating. It should be  contrast with our case when entire energy due to the annihilation will be released.  The numerical difference,  however, is a very minor effect as all   DM particles on NS surface become relativistic objects irrespective to the models. As a result, the difference in  equilibration   temperatures  on the surface does not lead to any  qualitative  observable effects in comparison with previous analysis. In particular,  instead of $T\approx 1750K$    from WIMP type models  \cite{PhysRevLett.119.131801}  we would get $T\sim 3000K$ from AQN model, if all released energy is thermalized. 

Now  we would like to make few comments on comparison of (\ref{total_power_NS}) with analogous estimates for the  solar corona  (\ref{total_power})  where the same annihilation events of the AQNs in solar atmosphere  generate fundamentally  new phenomenon representing the resolution of the solar corona heating puzzle within AQN framework  as explained in   previous sections \ref{nanoflares} and \ref{identification}.

The dramatic differences in  luminosities  (four orders in magnitude) between  (\ref{total_power_NS}) and  (\ref{total_power})  is related to the fact that   the impact  parameters are very different for this two cases. Indeed, factor $10^4$ between   (\ref{total_power})  and (\ref{total_power_NS})  can be understood as the ratio:
\be
\label{ratio}
\left(\frac{b^{\odot}_{\rm cap}}{ b^{\rm NS}_{\rm cap}} \right)^2\sim \left(\frac{10^6~\rm km}{10^4~\rm km}\right)^2\sim 10^4.
\ee  
As a result, the luminosity  ($\sim 10^{27} {\rm erg\cdot s^{-1}}$) radiated from corona     (\ref{total_power}), though  represents only $\sim 10^{-6}$ fraction of the total luminosity of the Sun, nevertheless produces the profound  observable effects in form of the EUV and X rays emission from corona.  In contrast, the   observation of the emission  (\ref{total_power_NS})  from NS is unlikely to be directly observed anytime soon.  

Indeed, the value for the surface temperature $T\approx 3\cdot 10^3 \rm K$ as estimated above  is way below of the present observational capabilities, and we shall not elaborate on this effect of heating due to the direct  AQN annihilations in NS atmosphere and the crust in the present work. Precisely this effect (heating of the very old stars due to  the direct energy injection by DM particles in form of WIMPs) was the main subject of the  analysis in most of previous studies \cite{PhysRevD.77.023006,PhysRevD.82.063531,PhysRevD.81.123521,PhysRevD.83.083512,
PhysRevD.85.023519,PhysRevD.87.123507,PhysRevD.89.015010, 
PhysRevLett.113.191301,PhysRevD.96.063002,PhysRevLett.119.131801,Bramante:2017ulk,Raj:2017wrv,Bramante:2023djs}. This is precisely the main conclusion of Sect.\ref{sect:DM} that DM accretion cannot play any role in heating of the NS to temperatures in the range $T\sim 10^5K$ as observed. This conclusion was entirely based on canonical assumption within the 40 year old paradigm that the DM is  represented by a   fundamental field   in form of a microscopical particle such as WIMP. 
 
In contrast, the AQN is a complex macroscopical object, outside of this  canonical paradigm. Therefore, it may play another role (as a trigger, see below) along with the  effect mentioned above.  

 This work is focused precisely on another consequence of the AQN framework when the nuggets play the role of the triggers which may ignite and initiate much larger events similar to the flares in the Sun as discussed in sections \ref{AQN-flares} and \ref{m_reconnection}. This effect is not shared by any other DM models when the DM particles are represented by   fundamental local quantum fields, such as WIMPs.  

When AQN serves as a triggers of a large event such as flare in the Sun,    the dominant portion of the energy   feeding such event  is  coming  from a strong magnetic field of NS (not from the  AQN itself)  converting  its energy into the radiation in broad frequency bands.  If this  happens  there could be  many  dramatic  observable effects, which precisely represents  the topic of the present studies.

\subsection{Mach number and shock waves}\label{sect:Mach1}
The goal of the present section is to argue that the AQNs can serve as the triggers which may initiate the magnetic reconnection
similar to our discussions in sections  in sections \ref{AQN-flares} and \ref{m_reconnection} in context of the    solar flare physics. 
Our arguments are based on estimation of different parameters such as $\beta$ and  the Mach number $M$
for NS environment. 
 
    In what follows it is   convenient to parametrize the velocity of an  AQN when it enters the NS atmosphere  in terms of the proper $\eta_\mu$ velocity and 4-momentum $p_\mu$ defined  in the usual way:
\be
 \label{proper}
\eta_\mu=\gamma(c,\vec{v}), ~~~  \gamma=\frac{1}{\sqrt{1-v^2/c^2}}, ~~~  p_{\mu}=M_N\cdot \eta_{\mu},
 \ee
  where $M_N\simeq m_pB$ is the AQN's rest mass expressed in terms of the proton mass as reviewed in section \ref{basics}.

  The key observation here is that  the Mach number $M\gg 1$ is always  very  large   for a typical AQNs entering the NS atmosphere,
\beq
 \label{Mach1}
 M &\equiv  &\frac{\sqrt{v^2_{\perp}+ v^2_{\parallel}}}{c_s} \\
&\simeq &  6\cdot 10^3\cdot  \left({ \frac{ {  10^5  K}}{{\rm }T}}\right)^{1/2}\cdot  \left(\frac{\rm 10^6~cm}{R_{\rm NS}}\right)^{1/2} \cdot \left(\frac{M_{\rm NS}}{2~M_{\odot}}\right)^{1/2} \nonumber 
 \eeq
is much larger than one. 
\exclude{
 One can also consider the Alfv\'{e}n Mach number $M_A$ defined in conventional way as:
 \be
 \label{Mach-alfven}
 M_A \equiv \frac{\sqrt{v^2_{\perp}+ v^2_{\parallel}}}{v_A},
 \ee
 where the Alfv\'{e}n velocity $v_A\simeq  c$
  is very close to the speed of light  in NS environment.  At the same time 
  the nugget's velocity $v_{AQN} $  is only a fraction of the speed of light.   Therefore $M_A< 1 $ in general.  In a former  case when $M \gg 1$ and $M_A<1$ one should expect the so-called ``slow shock wave", while  in a latter case when $M \gg 1$ and $M_A>1$, very close to the reconnection region, 
  one should expect the ``fast shock waves" \cite{Landau}.
 }  The shock wave may initiate a large event similar to solar flares  considered   in sections \ref{AQN-flares} and \ref{m_reconnection}.

    Another  important parameter which determines   importance of the magnetic pressure in comparison with kinetic pressure is  dimensionless  parameter $\beta$: 
 \beq
 \label{beta1}
 \beta\equiv \frac{8\pi  p}{{\cal{B}}^2} \sim 10^{-18}  \left(\frac{n}{10^{10} ~{\rm cm^{-3}}}\right) \left(\frac{T}{10^6 K}\right) \left(\frac{10^{10}~ G}{{\cal{B}}}\right)^2~~~~~
 \eeq
 where for numerical estimates we use typical parameters for NS. The crucial difference with the solar corona here is that $\beta\ll 1$ is  very small everywhere on the NS surface. It is not the case for the solar corona, where $\beta\ll 1$  only in active regions, while $\beta \gg 1 $   in quiet regions of the solar surface. As a result, the solar flares occur only in active regions with $\beta\ll 1$ when magnetic reconnection 
 could  in principle take place  as discussed in sections \ref{AQN-flares} and \ref{m_reconnection}, while in NS the magnetic reconnection could occur everywhere at any given moment, and may occupy entire NS surface as condition $\beta\ll 1$ holds everywhere. 
\exclude{
 Our comment here goes as follows.  In close vicinity of the reconnection region where the magnetic field strongly fluctuates the Alfv\'{e}n Mach number $M_A>1$ could be larger than one (locally), depending on specific  details, including the initial direction velocity of the nugget. Indeed, the magnetic field must change the direction  at the point of reconnection, and therefore it must vanish (locally) at some point. Therefore, the global characteristic determined by parameter  $\beta$ averaged over large volume as defined  by (\ref{beta1}) of the system does not reflect the local features of the system where $\beta$ could deviate from average value  (\ref{beta1}) due to the strong (local) fluctuations of the magnetic field in the vicinity of    (would be) the reconnection area where $\cal{B}$ is small and formula (\ref{beta1})  does not reflect the local dynamics. 
  }

One more parameter which characterizes the NS atmosphere   is the electron number density $n_e$ where the AQN-induced shock wave may propagate.  The uncompensated charge density of electrons and ions  in the NS atmosphere  is not vanishing  due to the so-called Goldreich-Julian (GJ) effect when the magnetic field spinning through a very good conductor produces the electric field which separates the charges. To be more precise the GJ number density is proportional to $n_{GJ}\propto {\mathbf \Omega}\cdot {\cal{B}}$. Numerically, it can be estimated as follows,  see e.g. \cite{Kumar:2017yiq}: 
\be
\label{n_e}
n_e\approx \frac{2{\cal{B}}}{ecP}\frac{R_{\rm NS}^3}{r^3}\sim \frac{10^{10}}{\rm cm^{3}}\left(\frac{1s}{P}\right)\left(\frac{{\cal{B}}}{10^{10}G}\right)\left(\frac{R_{\rm NS}^3}{r^3}\right),
\ee
where $P$ is the pulsar period, and $r$ distance from the centre of the star. 

 If we assume that magnetic reconnection indeed occurs as a result of the AQN   triggering   event,  what could be the 
  energy injection rate in this case? Is it sufficient to heat the old NS  to the high temperatures such as $T\sim (10^5-10^6)K$ as observed?
 The  corresponding   energy  is determined by the energy of magnetic field as a result of successful  reconnection. This injected  energy is dramatically larger in comparison with  the energy released  due to the direct annihilation events  of the AQNs in the NS atmosphere discussed in previous subsection  \ref{sect:AQN-energy}. The corresponding 
  estimates   will be presented  in the next section.

  \section{Magnetic Reconnection as the heating source of NS}\label{NS-heating} 
    We are now prepared to present  our order of magnitude estimates to argue that 
     the heat being   released   as a result of  the   AQN-induced magnetic reconnection events is sufficient
     to heat the old NS  to   high  temperature $T\sim 10^5K$ as observed.   
First, we estimate the total hitting rate $\dot{N}$ of NS by  AQNs.  It can be estimated by dividing  formulae   (\ref{rate}) to $M_{N}$, i.e.
 \be
\label{rate-AQN}
\frac{dN}{dt}\approx \pi b^2_{\infty}v_{\infty} \left(\frac{\rho_{\rm DM}}{M_{N}}\right)\approx \frac{10}{\rm s}
\left(\frac{10^{25}}{\langle B\rangle}\right) \cdot \left(\frac{M_{\rm NS}}{2~M_{\odot}}\right),
\ee  
 where  $\rho_{\rm DM}$ is the local DM density in the vicinity of the NS and $M_N\approx m_pB$ is the mass of the AQN, see table \ref{tab:basics}.
 
 Our next task is to estimate the minimal required energy   to heat the NS surface to the temperature $T\approx 10^5K$.
   The   required energy rate to be injected to heat the NS's surface has been estimated previously in (\ref{NS-emission}) and is given by 
 \be
 \label{NS-required}
  L=4\pi R^2\sigma T_s^4\approx   10^{29}\left(\frac{T_s}{10^5 \rm K}\right)^4\rm \frac{erg}{s}. 
 \ee

The  next task is to estimate the  total   available magnetic energy  above the NS's surface. After that one can    estimate the portion of the energy which should be converted to the heat to  equalize 
 the radiation  loss (\ref{NS-required}).  
 The total magnetic energy $E^{\rm tot}_{\rm mag}(A)$ above the NS's surface can be   estimated as follows
 \beq
 \label{m_energy}
 E^{\rm tot}_{\rm mag}\simeq\frac{1}{8\pi}  \int_{r\geq R_{\rm NS}}d^3x {\cal{B}}^2  
\simeq  10^{36}{\rm erg} \left(\frac{{\cal{B}}_{\rm surf}}{10^{10}~{\rm G}}\right)^2,  
 \eeq
 where we used a simple dipole formula ${\cal{B}}\simeq  {\cal{B}}_{\rm surf} {R_{\rm NS}^3}{r^{-3}}$  for the estimate\footnote{\label{B-dipole}One should mention here that the estimate (\ref{m_energy}) represents in fact a lower bound as the relevant magnetic field could be much stronger than a simple dipole formula would suggest, see also comments in Sect. \ref{sect:magnetic}.}.  
 
 To make further progress with our computations  we assume that every hit  by the AQN  of the NS triggers and initiates  a shock wave  which consequently generates  the magnetic reconnection. This assumption is very reasonable as a similar assumption 
 for solar flare  gives very reasonable estimates for the rate and strength of solar  flares as reviewed  in sections \ref{AQN-flares} and \ref{m_reconnection}. 
 Indeed, in both cases the relevant parameters $M\gg 1$ and $\beta\ll 1$ and the shock waves are very likely to be formed. The difference between the Sun and NS is, of course, that flares in the Sun can be generated in active regions only (which accounts for very tiny portion of the solar surface), while in NS the shock wave may develop anywhere on the  entirely NS's surface.  
 
 With this assumption we introduce parameter $\epsilon\ll1$ which describes  a small portion of the total magnetic energy (\ref{m_energy}) which will be converted to the heat after AQN struck the NS and  triggers the shock wave leading to the magnetic  reconnection with consequent heating the surface, i.e $E_{\rm heat}   \equiv \epsilon E^{\rm tot}_{\rm mag}$. The corresponding parameter $\epsilon$ is estimated from the following condition:
 \be
 \label{equilibration} 
 \epsilon E^{\rm tot}_{\rm mag}\cdot \frac{dN}{dt}= L \;\;\;\; \Rightarrow \;\;\;\; \epsilon\approx \frac{L}{\dot{N}E^{\rm tot}_{\rm mag}},
 \ee
 where parameters $\dot{N}$,  and $L$ are given by (\ref{rate-AQN}) and  (\ref{NS-required}) correspondingly. 
 Numerically, parameter $\epsilon$ can be estimated as follows:
 \be
 \label{epsilon}
\epsilon\equiv\frac{E_{\rm heat}}{E^{\rm tot}_{\rm mag}}, \;\; \; \epsilon\approx 10^{-8} \left(\frac{T_s}{10^5 \rm K}\right)^4  
 \left(\frac{10^{10}~{\rm G}}{{\cal{B}}_{\rm surf}}\right)^2 \left(\frac{\langle B\rangle}{10^{25}}\right),  
 \ee
 which implies that the total magnetic energy is more than sufficient to heat old NS to explain the puzzling observations reviewed in Sect.\ref{NS-heating}.
 In fact,  only very tiny portion of the magnetic energy ($\sim 10^{-8}$) needs to be converted to the heat at each event of the magnetic reconnection triggered by the AQN. In fact the numerical value for $\epsilon$ is expected  to be even smaller because the relevant magnetic field is likely  to be much stronger than a simple 
 dipole formula would suggest, see footnote \ref{B-dipole}. In what follows we shall argue that   this condition  can be indeed naturally satisfied.  
 
 At this point one could wonder what went wrong with  the old (and naively, very generic) argument from Sect. \ref{NS-heating-mechanisms},  suggesting that magnetic field cannot play any role in heating of old NS according to (\ref{B-decay})? The answer is  related to  two    new elements which were completely missed in  naive estimate (\ref{B-decay}). First, the magnetic field locally could be much stronger than a simple dipole formula would suggest as we already mentioned in footnote \ref{B-dipole}. Furthermore, as we argue below in Sect. \ref{sect:reconnection} the energy which is powering the magnetic reconnection is related to the magnetic helicity $\cal{H}$, see  Appendix \ref{sect:helicity} for definition and basic features of the magnetic helicity $\cal{H}$. 
 Another novel element is  the relevant time scale which enters (\ref{rate-AQN}) and which is dramatically different from time scale entering the naive estimate (\ref{B-decay}).   This portion of the magnetic energy  could  be quickly restored after every AQN-induced event of reconnection  with rate  (\ref{rate-AQN}). A possible mechanism for such  ``refill" of the magnetic helicity is discussed in items 10, 11 in next Sect. \ref{sect:reconnection} and in Appendix \ref{sect:helicity}.

  \section{Magnetic Reconnection in NS. Basic ingredients}\label{sect:reconnection} 
  In this section we want to formulate the basic ingredients of the proposal   supporting our main result formulated in previous section. It 
  suggests that the magnetic reconnection triggered by the AQN may indeed heat the old NS. In this sense we suggest an alternative mechanism,    which we claim, is capable to generate  enough energy to heat the old NSs to explain the puzzling  observations listed in Sect. \ref{NS-heating-mechanisms}.
  
  First of all, we would like to notice that the computation of the parameter $\epsilon$ defined by (\ref{epsilon}) from the first principles is not feasible at this point due to very complicated dynamics of the strongly coupled system AQN-NS. In particular, it includes the  evolution of the shock waves, developing of the  turbulence and many other accompanied phenomena  when the energy transfer occurs from a body moving with 
  enormous Mach number as estimated by (\ref{Mach1}). 
  
  Nevertheless, there are many systems where the magnetic reconnection is known to occur and is believed to   power very energetic events such as solar flares discussed in previous  sections \ref{AQN-flares} and \ref{m_reconnection}. One can use the corresponding observations to 
   support or refute some of the assumptions on dynamical features of the magnetic reconnection in NS
   based on experience with the solar flare events. 
  
  Another system where magnetic reconnection is believed to play a crucial role (see e.g. \cite{Kumar:2017yiq}) is the   magnetars  where the so-called Fast Radio Bursts (FRB) are erupted as a result of the magnetic reconnection, see recent review on the topic \cite{Zhang:2022uzl}.   While the idea that FRBs are powered by the magnetic field   transferring  an enormous energy 
  to the radio emission is commonly accepted in the community, the suggested triggering mechanisms which could initiate the magnetic reconnection dramatically vary: from a crust cracking at the NS's surface to sudden triggers from an external event, see \cite{Zhang:2022uzl} for review. 
  
  Our proposal \cite{vanWaerbeke:2018nyj} that the DM particles in form of the AQNs play the role of the triggers for FRBs is exactly from the last category when an external object initiates the FRB. In our case the external object is the AQN.  In many respects our present proposal that the AQNs could  be the triggers of the magnetic reconnections in old  NS which may heat the NS's surface to explain  the puzzling observations as reviewed in Sect. \ref{NS-heating},
  is  very similar to proposal \cite{vanWaerbeke:2018nyj} in context of FRBs.  
  
  There are many quantitative differences between these two cases, of course: the magnetars from proposal \cite{vanWaerbeke:2018nyj} are much younger, have much stronger magnetic field, have much higher surface temperature than the old NS which is the topic of the present work. However, the basic fundamental concept in these two cases is the same: the AQNs can serve as the triggers to ignite and initiate the magnetic reconnection which may feed the very energetic explosions in both cases. 
  
  The two systems mentioned above (flares in solar corona and FRB in magnetars) will be considered as 
 a  tool box which allows us to test the main  proposal of the present work. Indeed, by comparing   one or another  assumption from the present proposal with similar (analogous) studies of the flares in solar corona or  FRB in magnetars one can support or refute a corresponding assumption. This is precisely the approach we are  advocating   in this section. 
 
 We present below the basic elements of our proposal, item by item. In many items  we explicitly point out some similarities between our present system and previously considered systems --flares in solar corona and FRB in magnetars. Therefore, our assumptions can be confronted with available observations. 
 
 1. The basic conditions such as $M\gg 1$ and $\beta\ll 1$ are satisfied according to (\ref{Mach1}) and (\ref{beta1}), similar to analysis in active regions in the Sun and FRB; 
 
 2. As a result of these conditions the AQN may serve as the trigger  to initiate the magnetic reconnection
 as environment in all cases is also very similar. Indeed,  the density of highly ionized plasma  estimated as (\ref{n_e}) which is close to the density of  the solar corona;
 
 3. Therefore, we expect a strong shock wave generated by   propagating AQN, such that the  pressure $p$ and  temperature $T$  
 experience strong    discontinuities  according to (\ref{shock1}). This very strong impulse   in very small and very localized area (determined by the AQN's path) for very short period of time  $\tau$ may lead to a successful   magnetic reconnection;
 
 4. A typical time scale $\tau$ where the shock wave develops is determined by the velocity $v_{\rm AQN}$ of the AQN in vicinity of the NS surface, which is close to $c$. Therefore 
 \be
 \label{tau}
 \tau\sim \frac{d}{c}\sim \frac{10  \rm ~km}{c}\sim 0.3\cdot 10^{-4} \rm s , 
 \ee
  where distance $d\sim  10 {\rm ~km} $ is determined by the region where the density of the ionized gas is sufficiently large (\ref{n_e});
 
 5. The time scale $\tau$ determines the size of  the cone where the shock wave develops and where the pressure $p$ and  temperature $T$  experience strong    discontinuities, similar to the solar flare analysis (\ref{initial1});
 
 6. This AQN-induced shock will trigger the magnetic reconnection  in the area $A$ estimated as 
 \beq 
 \label{A}
 A\approx d c_s \tau \approx \frac{d^2}{M}\sim 10^{-4} d^2,
 \eeq 
 which represents a small portion $\sim (10^{-4}-10^{-5})$ of the NS's surface.  This area has the same physical meaning as estimate (\ref{initial}) for the solar flare which defines the area for  initial stage of the magnetic reconnection. 
   The difference with the solar flare is that the magnetic reconnection could only occur in a small active region of the Sun where magnetic field is large and condition $\beta\ll 1$ is satisfied, while the reconnection  in NS may occur anywhere  on the  surface as condition $\beta\ll 1$ is satisfied everywhere;
 
   7. Therefore,  if magnetic reconnection starts in  one location   it may quickly sweep out (potentially) entire NS's surface.  In this case  the total energy of the event in NS ({\it second stage} in classification (\ref{developed}) in context of solar flares)  represents a finite portion of   the integral (\ref{m_energy}). It must be contrasted  with   the solar flare  estimate  (\ref{developed}) which  represents a very tiny  portion of the surface in comparison with total surface of the Sun (the so-called active regions, the sunspots);
   
   8.   The speed of magnetic reconnection is governed by dimensionless parameter    $\beta_{\rm in}$ in notations \cite{Kumar:2017yiq} and \cite{vanWaerbeke:2018nyj},  where ``in" in  $\beta_{\rm in}$ stands for inflow speed. It 
   must be sufficiently high for fast successful reconnection, but it must be much slower than the speed of light  $\beta_{\rm in}\ll 1$. If we formulate this condition in terms of the reconnection typical time scale $\tau_{\rm in}\equiv d \cdot \beta^{-1}_{\rm in}$ the following hierarchy scales must be satisfied\footnote{\label{times}Parameter $\beta_{\rm in}$ was estimated in \cite{Kumar:2017yiq} in context of FRB physics. It determines the duration of the magnetic reconnection as $\tau_{\rm FRB}\propto \beta_{\rm in}^{-1}\approx 10^{-3}$ s. Parameter $\tau_{\rm FRB}$ plays the same role as $\tau_{\rm in}$ in (\ref{hierarchy}). Similar parameter (duration  of the large flare) in context of the solar physics could be as long as few hours, while the   duration of the initial stage of flare  lasts about 10s, see  (\ref{initial}). In all  cases the ratio between duration of the trigger event (due to the AQN)  and the magnetic reconnection itself is about $\sim 1\%$.}:
   \be
   \label{hierarchy}
   \tau \ll \tau_{\rm in}\ll \dot{N}^{-1} \;\;\; \rightarrow \;\;\; \rm 0.3\cdot 10^{-4} s\ll \tau_{\rm in}\ll 0.1 s,
   \ee
   where parameter $\tau$ is determined by the AQN   serving as a trigger, see (\ref{tau}), while  time scale $\dot{N}^{-1}$ is a typical time scale between two independent consecutive events according to (\ref{rate-AQN});
   
 9. The dynamics of magnetic reconnection   studied  in context of FRB suggests that the energy powering FRB is the magnetic helicity ${\cal{H}}$, see Appendix \ref{sect:helicity} for the definition and  short introduction into the subject.  We assume that this feature on the dominant role of the ${\cal{H}}$  holds   for the present proposal as well.  The basic argument for this assumption is that    the environments of NS and magnetars are very similar.  
 
 The magnetic reconnection implies that the electric field $E_{\parallel}$  during the reconnection time $\tau_{\rm in}$ will  be induced, see  \cite{Kumar:2017yiq,vanWaerbeke:2018nyj}. Its direction should be parallel  to the original static magnetic field $\cal{B}$  with the coefficient to be proportional to $\beta_{\rm in}$, i.e.
 \be
 \label{E}
 E_{\parallel} (\rm induced)\propto \beta_{\rm in}\cal{B}.
 \ee
 The condition (\ref{E}) unambiguously   implies  that  very specific $E\&M$ configuration with     $ \sim \vec{E}\cdot \vec{{\cal{B}}}\propto  \beta_{\rm in}{\cal{B}}^2$  must be generated during the reconnection. The main feature  of this configuration is that it enters the formula (\ref{time-derivative}) which describes the dissipation of the 
 magnetic helicity ${\cal{H}}$. Precisely the dissipation of the 
 magnetic helicity ${\cal{H}}$ powers the magnetic reconnection. Assuming $\beta_{\rm in}\sim 1\%$ 
 (which is within the window (\ref{hierarchy})) one can infer that every event of the magnetic reconnection 
 will convert $\sim 1\%$ of its magnetic helicity   into the heat along with 
  other  radiation losses such as X-rays.
 Such relatively high efficiency rate is obviously more than sufficient to generate energy   heating  the NS surface according to (\ref{epsilon}) even if one assumes that only a small finite  portion of the surface (rather than entire NS)  will be experiencing the magnetic reconnection. 
 
  10. It is very likely  that there are some mechanisms which restore the   energy associated with magnetic helicity    dissipation (due to the magnetic reconnection) and restore its equilibrium value. Indeed, there are some observations in FRB context suggesting that  frequency of some  FRB repeaters is enormously high. For example, rFRB 20121102A
  emitted a total amount of energy $\sim 3.4\cdot 10^{41} \rm erg$ in the radio band from 1652 bursts detected in 59.5 hours in a 47 day time span  \cite{Li:2021hpl,Zhang:2022uzl}. Assuming that this activity represents a typical behaviour of the FRB repeaters one could infer that there must be a mechanism which restores the  magnetic source of the energy\footnote{The total energy emitted would exceed $6.4\cdot 10^{45}\rm erg$ assuming a radio efficiency $10^{-4}$, which we consider is already too high. This is a substantial fraction of the available magnetic energy from a magnetar  \cite{Zhang:2022uzl}. This estimate again strongly suggests that there must be a mechanism   restoring the  magnetic  energy after the eruptions as example of the rFRB 20121102A already poses significant energy budget issues \cite{Zhang:2022uzl}.}.  
  
  Based on this   observation (in FRB context) we assume that  there should be an operational   mechanism
    which restores the energy reservoir  (\ref{m_energy}) by equilibrating the system   after the  events of reconnection which heat the NS's surface, see also footnote \ref{B-dipole} on numerical value for relevant value of $\cal{B}$ entering formula (\ref{m_energy});
    
\exclude{
11. Another related question which emerges from the above estimates   is as follows: what is the time scale for   the magnetic helicity 
 ${\cal{H}}$ to restore its equilibrium value and its initial magnetic field which existed before a  reconnection event? 
  }
11. A possible  mechanism of equilibration which could potentially restore the magnetic energy 
had been discussed previously in a very different context in \cite{Charbonneau:2009ax}. Important  conclusion of these studies was that 
 the magnetic helicity ${\cal{H}}$ which is powering the magnetic reconnection (see item 9)  is a very generic feature of the NS, and in fact there are many observational evidences  suggesting that  the magnetic helicity ${\cal{H}}$ must be  present in NS. 
 We overview the basic results of these studies  below and refer for the details to the Appendix  \ref{sect:helicity};
 
 12. The basic argument presented in \cite{Charbonneau:2009ax}
 is  that the so-called topological non-dissipating currents 
 will be induced in the NS  as a result of quantum anomalies   in high density QCD. These currents might be responsible for many observed phenomena such as NS kicks,  toroidal magnetic field, the magnetic helicity, to name just a few.  
 This is very generic and  well known phenomenon  in QCD. It is  related to the asymmetry between left handed and right handed chemical potentials which could be generated by    $\cal{P}$-odd weak interactions. Formally, it could be expressed as    generation of  the so-called axial chemical potential $\mu_5\equiv1/2(\mu_R-\mu_L)$.
  
 The $\mu_5\neq 0$ in context of the  heavy ion physics in QCD   leads to a number of $\cal{P}$   odd effects, such as chiral magnetic effect, chiral vortical effect, charge separation effect, to name just a few.   
 This field  of research initiated in \cite{Kharzeev:2007tn}  became a hot topic  in recent years as a result of many interesting theoretical and experimental advances,    see recent review papers \cite{Kharzeev:2009fn,Kharzeev:2015znc} on the subject.  
 
 In context of the present work of NS physics, one should mention that there is a  strong observational evidence, see e.g.\cite{Page:2007br} and references therein, supporting the presence of the toroidal magnetic field which unambiguously suggests that the magnetic helicity $\mathcal{H}$ must be non-zero in neutron stars.  We consider this as an indirect observational evidence supporting  claim that $\cal{P}$-odd topological currents  \cite{Charbonneau:2009ax}
 had been induced at some early  moment  in the star's evolution.  
 
 13. If one assumes   that  the magnetic helicity $\mathcal{H}$ is present in the system at the moment of magnetic reconnection, it is   expected from (\ref{E}) that the energy feeding the magnetic reconnection  comes from the magnetic helicity ${\cal{H}}$, which is directly related to the toroidal magnetic field in the system, see Appendix  \ref{sect:helicity} for  additional comments.
 One should emphasize that the toroidal magnetic field is generated by non-dissipating topological current (\ref{top}), in contrast with typical  dipole type field  (\ref{m_energy}) which is generated by usual dissipating currents  satisfying the  conventional Ohm's law (\ref{ohm}) in the core of the NS. 
   
  14. In order to understand what happens when the magnetic reconnection event occurs, one should recall that the NS system (which includes  the dipole type magnetic field, the toroidal magnetic field, the magnetic helicity ${\cal{H}}$, and the non-dissipating topological current with many other conventional components) is a very  complicated dynamical self-interacting and self-equilibrating  system.   It implies that if some elements of the system suddenly get changed the other elements of the system will adjust their values     to restore the equilibrium of the system. 
    
  In context of the reconnection events it implies that   the removing (due to reconnection) some value of the magnetic helicity ${\cal{H}}$ will result in modification of the currents and chemical potential $\mu_5$ to restore its equilibrium values. This equilibration is formally expressed by (\ref{total}). 

 The reservoir of the chemical potential $\mu_5$ is truly  enormous as estimate (\ref{reservoir}) suggests.    Therefore, we propose  that the mechanism of equilibrating the magnetic helicity $\cal{H}$ as suggested above, in principle is capable to    restore the energy associated with   magnetic helicity after the events of reconnection, see Appendix  \ref{sect:helicity} with additional comments and clarifications.  
 
 We conclude this section with the following comment. All analytical expressions  as presented above should be taken very cautiously with the grain of salt as they had been derived  from comparison with a  different system with dramatically different parameters. Nevertheless, the basic fundamental principles (the magnetic reconnection triggered by an external object) are very much the same. As a result we expect that our estimates give a  qualitatively correct big picture. 
 The magnetic reconnection, its evolution, and the triggering mechanisms are obviously the prerogative of numerical simulations which can support or refute  the 
  hypothesis advocated in this work. Therefore, we strongly advocate the researchers in relevant fields to consider this picture seriously. 
  A hope is that the recent advancements in the field can successfully attack these complicated   technical problems and test the heating mechanism as advocated in this proposal.

Another option to test this proposal (which is  complimentary to   numerical tests mentioned above) is to    measure  some specific  observables which always accompany the magnetic reconnection.  This is the topic for the next section.

   \section{X-rays as  the  indicators of the magnetic reconnection in NS}\label{x-ray}
  First, we start by mentioning that  it has been known for quite some time that the solar flares are normally accompanied by strong X ray radiation, see
   e.g.   \cite{Bruevich_2017} for review of the last complete solar cycle No -24.    Furthermore, the X-rays are considered to be  a good indicator for large flares because the X ray intensity dramatically increases few moments before a flare starts. 
    
    In addition to that it  has been point out in   \cite{Bruevich_2017} that the solar pre-flare enhancement in form of the X rays propagates from higher levels of corona into lower   corona and chromosphere, see Fig. 8 in  \cite{Bruevich_2017}. 
   This pre-flare enhancement  is very  puzzling and unexpected phenomenon as pre-flare propagates from top to bottom. Nevertheless, this 
    unusual temporal  and spatial  patterns of propagation    have very natural explanation within AQN framework (reviewed in  Sect. \ref{m_reconnection} in context of the solar flares) because  the AQNs propagate (and ignite  the magnetic reconnection) from outer space to the surface. 
    
  Now we return to our main topic of the NS. The leitmotiv   of the proposal advocated in this work is that   two naively very different phenomena (in dramatically different systems): \\
    1. the heating mechanism  of relatively old NS\\
    2. the solar flares\\
    are in fact very similar as they both powered by the same mechanism of  the magnetic reconnection (according to the proposal) triggered and initiated by the dark matter AQN particles. 
       Therefore, we expect that  a number  of accompanying effects associated with magnetic reconnections must   also manifest themselves in very similar ways in both cases. One of such profound accompanying  effect of  large solar flares is the X-ray emission, see first paragraph of this section. Therefore, we expect that a similar X-ray emission must be also present in NS if the    NS heating is indeed powered by magnetic reconnection as advocated in this work. 
       
     The main   goal of this section is to estimate the X ray intensity from  NS   by using  (for normalization)    the observed and well recorded intensity of the X-ray radiation during the solar flares.   In this estimate   we use the same logical steps   as in previous sections \ref{NS}- \ref{sect:reconnection} by  comparing one and the same phenomenon  in two different systems. 
     
     \subsection{X rays from solar flare as normalization point}\label{sect:X-ray}
            The starting point of our estimates is as follows. The peak of the X rays in the   band $0.1-0.8~ \rm nm$  of the flare of August 9. 2011  is recorded as follows, see Fig 8 in   ref.\cite{Bruevich_2017}:
            \beq
            \label{F_sun}
            F^{\odot}_{X}&\approx&    10^{-3}\rm \frac{W}{m^2}\approx  \frac{erg}{s\cdot cm^2},\\
              E &\in& (1.5-12.4)~\rm  keV, \nonumber
            \eeq
            while the average $\la F^{\odot}_{\rm X}\ra$ of the X ray emission during the flare can be estimated as  $ \la F^{\odot}_{\rm X}\ra \approx  10^{-2}~ \rm erg\cdot s^{-1}  \cdot cm^{-2}$ which is almost two orders of magnitude smaller than the peak value  (\ref{F_sun}).
      At the same time the total flux from  the flare emitted from all frequency bands can be estimated as 
       \beq
            \label{ratio1}
            F^{\odot}_{\rm tot}&\approx& 10 \left(\frac{E_{\rm tot}}{10^{32} \rm erg}\right){\rm \frac{erg}{s\cdot cm^2}},  \nonumber \\ \xi &\equiv&\frac{ \la F^{\odot}_{\rm X}\ra}{F^{\odot}_{\rm tot}}\approx (10^{-2}-10^{-3})
            \eeq
        assuming that the flare   lasts about $ \rm 1h$ and the total released energy $E_{\rm tot}$ during a large  flare was between  $10^{31} \rm erg$ and $10^{32} \rm erg$.
        The order of magnitude  estimate (\ref{ratio1}) suggests  that the X ray flux   represents  less than  1\% portion of a large flare.  We use this  ratio $\xi$ to estimate the X ray flux from NS due to the magnetic reconnection in next subsection.

        \subsection{X rays from NS originated from  magnetic reconnection}\label{sect:X-ray-NS}
        In our estimates which follows we assume that the observed temperatures of sufficiently old  NS are entirely 
        saturated by the magnetic reconnection mechanism as advocated in this work. This is obviously a  strong assumption.
        However, as reviewed in Sect. \ref{NS-heating-mechanisms} any common  mechanisms  such as   roto-chemical heating 
        cannot explain many  observations, see Sect.\ref{sect:rotochemical}. In particular, the so-called Magnificent Seven stars cannot be explained by this mechanism with reasonable  changes of the parameters, see Fig. 3 in \cite{Yanagi:2019vrr}. There are many similar cases when the observed temperature  of NS dramatically exceeds 
the theoretical estimates and inconsistent with canonical picture of cooling, including additional mechanisms reviewed in Sect. \ref{NS-heating-mechanisms}. 

With this assumption in mind and assuming the black-body radiation spectrum we estimate the total flux of E\&M radiation 
(powered by  magnetic reconnections, as presented in previous Sect. \ref{sect:reconnection})  
as follows:
\beq
\label{F_NS}
 F^{\rm NS}_{\rm tot}&=& \frac{L}{4\pi r^2}\sim \frac{7\cdot 10^{32}}{4\pi r^2} \left(\frac{T_s}{10^6 \rm K}\right)^4 \rm \frac{erg}{s} \\&\sim&6\cdot 10^{-10} \left(\frac{T_s}{10^6 \rm K}\right)^4 \left(\frac{0.1 \rm  kpc}{r}\right)^2{\rm \frac{erg}{s\cdot cm^2}}\nonumber
\eeq
where $r$ is the distance to NS, while the luminosity $L$ is estimated in (\ref{NS-emission}).   

We need two more elements to complete our estimate for $\la F^{\rm NS}_{\rm X}\ra$ analogous to $\la F^{\odot}_{\rm X}\ra $
entering (\ref{ratio1}). First, we assume that the ratio $\xi\sim 10^{-3}$ defined by (\ref{ratio1}) is the same for the solar flares and NS because (according to our proposal)  two different phenomena are originated from the same physics of the magnetic reconnection as argued above. Another element which also important for our estimate of $\la F^{\rm NS}_{\rm X}\ra$ is the 
ratio of two different time scales: first,   the $\tau_{\rm in}$ is   reconnection time scale, which represents  the duration of the magnetic reconnection, while    $\dot{N}^{-1}$ is a typical time scale between two independent consecutive events as defined in (\ref{hierarchy}). As a result, we arrive to the following estimate for the X ray flux from NS due to the magnetic reconnection\footnote{The time scale $\tau_{\rm in}$ enters the formula (\ref{X_ray}) because X rays can be emitted exclusively during the reconnection period according to our proposal, similar to discussions of the X ray radiation during the solar flare from Sect.\ref{sect:X-ray}. At the same time the $\tau_{\rm in}$ is normalized by a typical time scale $\dot{N}^{-1}$, which represents a time scale before a next AQN hits the NS when  new reconnection starts and  new portion of the energy is injected into the NS's atmosphere. We assume that  the  energy  will be eventually thermalized, which justifies our formula for luminosity (\ref{F_NS}) with black body radiation spectrum.}: 
\beq
\label{X_ray}
\la F^{\rm NS}_{\rm X}\ra&\sim& \xi \cdot  \left(\frac{\tau_{\rm in}}{\dot{N}^{-1}} \right)\cdot F^{\rm NS}_{\rm tot}\sim (10^{-4}-10^{-5}) F^{\rm NS}_{\rm tot} ~~~ \\
&\sim&   (10^{-13}-10^{-14}) \left(\frac{T_s}{10^6 \rm K}\right)^4 \left(\frac{0.1 \rm  kpc}{r}\right)^2{\rm \frac{erg}{s\cdot cm^2}},\nonumber
\eeq
 where for the numerical estimate we use $\tau_{\rm in}\sim 10^{-3}\rm s $ (motivated by the FRB analysis) and  $\dot{N}^{-1}\simeq 0.1 \rm s$, see footnote \ref{times} with some comments.   
 
 One should emphasize that this is really an order of magnitude estimate- it is very hard to improve it as the estimate (\ref{X_ray}) includes large number of elements with unknown physics.  Furthermore, many NS are known to be emitters of the  X rays due to many other reasons (for example, due to the conversion of the spin-down power into the X rays). Therefore,  
 our estimation (\ref{X_ray})  should be considered as  an additional  contribution to the X ray emission. It obviously implies that it is very hard to discriminate the X ray emission flux as given by  (\ref{X_ray}) from many other canonical astrophysical mechanisms. However,  there are known  special   cases when conventional astrophysical mechanisms  produce very tiny contribution to the X ray radiation, in which case (\ref{X_ray})  could play the dominant role, and in principle could be discriminated from
 other mechanisms, see example below.
 
 It is instructive to present a numerical value of  the flux $\la F^{\rm J1856}_{\rm X}\ra$ for $J1856$  from Magnificent Seven (M7) stars which is the closest to Earth.  This NS belongs to the category  when the star's  temperature greatly exceeds  an  anticipated value as reviewed in Sect. \ref{sect:rotochemical}.   The corresponding  flux $\la F^{\rm J1856}_{\rm X}\ra$   is estimated from (\ref{X_ray}) as:
 \be
 \label{J1856}
 \la F^{\rm J1856}_{\rm X}\ra \approx (4\cdot 10^{-15}-4\cdot 10^{-16})  \rm \frac{erg}{s\cdot cm^2}, [\rm estimation]
 \ee
          where we use $r\approx 0.123~ \rm kpc$ and $T_s\approx 0.5\cdot 10^6 K $ for the numerical estimates. 
          It is   quite remarkable that the X ray emission has indeed been observed \cite{Dessert:2019dos} from this star 
         with $5\sigma$ excess in $(2-8) \rm~keV$ energy range with result:
     \beq
     \label{observations}
 \la F^{\rm J1856}_{\rm X}\ra = (1.5^{+0.7}_{-0.6})\cdot 10^{-15}  \rm \frac{erg}{s\cdot cm^2}, \;\;\; [\rm observations].\;\;\;
 \eeq    
 According to \cite{Dessert:2019dos} it is very hard to explain this excess of  X ray radiation by any conventional astrophysical sources or any  systematic effects.
                  This observation nicely falls into  the  interval (\ref{J1856}). We consider this result   as a highly nontrivial consistency check for the application of the AQN framework   to NS heating problem as the parameters entering the estimate (\ref{X_ray}) are based on dramatically different physics describing enormous range of scales in drastically  different contexts, 
                  from DM distribution to solar physics. All the corresponding parameters had been fixed long ago in application to different systems, without any attempt to modify or fit them to match the present observations. 
                     Therefore, (\ref{X_ray}) could be potentially many orders of magnitude off from the observed value (\ref{observations}). 
                     
                     In many aspects the similarity of the numerical values between
            (\ref{J1856}) and (\ref{observations})       is analogous to similarity between   the observed solar corona  luminosity $L_{\rm corona} \sim 10^{27} {\rm erg~ s^{-1}}$ in EUV and the AQN-induced luminosity  (\ref{total_power}) which is entirely determined by the DM parameters, not related to solar corona. 
  Therefore, we think    it is very hard to interpret the numerical agreement between     (\ref{J1856}) and (\ref{observations}) as simply an       ``accidental numerical coincidence". We think it should be interpreted, similar to mysterious solar corona EUV radiation,      as a result of some deep roots and connections between DM physics and NS physics, which is naturally incorporated by the AQN framework.

 
\section {Concluding comments and Future Developments} \label{conclusion} 
 The presence of the {\it antimatter} nuggets
 in the system implies, as reviewed in Sect.\ref{AQN}, that there will be  annihilation events (and continues injection of energy at different frequency bands) leading to  large number of observable effects on different scales: from Early Universe to the galactic scales to the Sun and the  terrestrial rare events. 

 In the present  work we focus on manifestations of these annihilation events on  physics of the NS. We proposed that DM in form of the AQNs may serve as the triggers  igniting the  large explosive events powered by the magnetic reconnection. The released energy as a result of these events may serve as the  heater  of NS as suggested in Sect. \ref{NS-heating}.
 This is precisely the additional source of energy which may resolve the   mysterious and puzzling observations  as reviewed in Sect. \ref{NS-heating-mechanisms} when the NS temperature is inconsistent with canonical cooling mechanisms. 
 
 One should emphasize once again that  a precise ``measuring" of the NS's surface temperature  is a very subtle point such that all recorded values  should be taken with   some scepticism,  see footnote \ref{subtleties} with a comment. Nevertheless, we believe that the observed discrepancy between ``measured" and predicted temperatures is a real physical effect, and we propose a specific mechanism which could be responsible for excess of the heating.  
 
 We do not need to repeat the key elements on  physics of the magnetic reconnection triggered by the AQNs as  presented in Sect. \ref{sect:reconnection} in this Conclusion. 
  Instead,  we want to mention below several phenomena which should accompany the proposed mechanism of the excess of  heating  of old  NS. As such these additional emissions  should be considered as   possible tests  and predictions for proposed mechanism of heating.  
 
   In Sect. \ref{sect:tests} we list some possible new tests which can substantiate or refute our proposal.    Finally,   in Sect. \ref{sect:paradigm} we describe  several  other mysterious and puzzling observations (outside  the NS system),  which can be understood within the same AQN framework. We consider this as     indirect support  for our proposal as the computations are based on  the same set of parameters of the AQN model reviewed in Sect. \ref{AQN}. 
 
    \subsection{Possible tests of the proposal}\label{sect:tests}
    As mentioned at the very end of  Sect. \ref{sect:reconnection} there are several tests which can substantiate or refute this proposal on heating mechanism of the old NS.
    One of them is a   numerical study of   evolution of the magnetic reconnection, its evolution, the triggering mechanisms which  is obviously  the prerogative of numerical simulations.  Another, complimentary approach  (which  represents the  topic of the present section)  is analysis of the 
    radiation in very broad frequency bands (from radio to hard X rays, and likely to gamma rays) which always accompany the heating mechanism suggested in sections  \ref{NS-heating} and  \ref{sect:reconnection}.
    
    There is enormous energy reservoir (\ref{reservoir}) which could  be converted to the magnetic helicity    and eventually to the heat and $E\&M$ radiation.   
    We specifically focus on hard X ray radiation in  Sect. \ref{x-ray} because we consider this frequency band is  the  most promising channel where this heating mechanism can be directly tested. Another possible 
    radiation  in the  radio frequency bands  which also accompanies the magnetic reconnection is expected to be less promising as the estimates in  Appendix \ref{radio} imply. 
    
    Essentially we suggest to study the hard X ray emission from other M7 stars as estimation (\ref{X_ray}) applies  to all of them.
   As we mentioned in Sect. \ref{NS-heating-mechanisms} all M7 stars have the temperatures which greatly exceed the expectations. 
      We interpret this inconsistency as the presence of the additional heating mechanism in form of the magnetic reconnection for all M7 stars
      in spite of subtleties related to  the ``measured" surface temperature as 
      mentioned in footnote \ref{subtleties}. 
      Therefore we predict that all M7 stars should emit the hard X rays which always accompany the reconnection  with flux being estimated in (\ref{X_ray}).   As mentioned in \cite{Dessert:2019dos} the observation of the hard X ray   from  $J1856$       
 with $5\sigma$ excess  (and not observations of a similar signal in other M7 stars) could be related to the fact that $J1856$ has  the most exposure time across all of the X-ray cameras that were   considered in \cite{Dessert:2019dos}. 
       
       Another possible class of NS where proposed heating mechanism could manifest itself is represented by  vey old pulsars. As mentioned in Sect. \ref{NS-heating-mechanisms} the  observed temperatures (well exceeding $T_s\gtrsim 10^5$ K) of many old pulsars  with  $t\gtrsim 10^8 \rm yr$ cannot be   explained by conventional   mechanisms, see review  \cite{Yanagi:2019vrr} with details. At the same time, there are many cases where such high  (and even much higher)  surface temperatures  have been observed. As an illustrative sample from this class  one could   consider PSR J0108-1431which  is a nearby ($r\simeq 0.13~ \rm kpc$), 170 Myr old pulsar. Its surface temperature is known to be very high: $ (1-5)\cdot 10^5 K$ 
       and it is hard to explain even with additional  heating mechanisms  \cite{Yanagi:2019vrr}. 
       
       It is also interesting to note that 
       this pulsar is observed in X ray with flux  $\la F^{\rm J0108}_{\rm X}\ra = (9\pm2)\cdot 10^{-15}  \rm \frac{erg}{s\cdot cm^2}$ in the $(0.3-8)~\rm keV$ band.  It is very similar in value to (\ref{observations}) observed in $(2-8)~\rm keV$ band,
        if one excludes the soft X ray segment from $\la F^{\rm J0108}_{\rm X}\ra $   representing  its dominant portion   as the corresponding spectrum has a power law with index $\gamma\simeq 2.2$, see 
  \cite{Pavlov:2008ng}.     This similarity between two different cases is consistent with our formula  (\ref{X_ray}) as the temperatures and distances for both NS are almost the same. 
  
  One should note that in the original paper \cite{Pavlov:2008ng}
  the relatively large   flux  $\la F^{\rm J0108}_{\rm X}\ra$ was entirely attributed to spin-down mechanism with enormously high X ray efficiency $\eta_X\simeq 0.4\cdot 10^{-2}$, while  for typical younger pulsars similar X ray efficiency is  two orders of magnitude smaller: $\eta_X\simeq(10^{-5}-10^{-4})$. In our view it is very hard to  justify such dramatic increase of the X ray efficiency for older pulsars from the theoretical side. We more incline to interpret a sufficiently high  X ray flux $\la F^{\rm J0108}_{\rm X}\ra$
  as a manifestation of the magnetic reconnection which   powers  the heating of this old pulsar, and accompanied to this heating the X ray emission as estimated in Sect.\ref{x-ray}. 
  
  Another possible test of the proposed heating mechanism is a study of   thermal pattern  on the NS's surface. The main observation here is that the magnetic reconnection is powered by magnetic helicity $\cal{H}$ (on large and small scales). Consequently, the toroidal magnetic field  is expected to play an essential role in the dynamics and heating of the NS as discussed in Sect. \ref{sect:reconnection}. It implies that the thermal  pattern on the NS's surface must be very different from canonical  poloidal dipolar magnetic field  (in which case   the  cold  region is always  localized along the equator while the  hot regions are always localized around  the poles). These topics are  obviously prerogative of numerical simulations which can support or refute  the hypothesis advocated in this work.  It  can be only accomplished with comprehensive numerical simulations of magneto-thermal coupled evolution which includes such elements as the magnetic reconnection, its evolution, and the triggering mechanisms. 
  
  In addition,   an observation of  the magnetic field during the solar flares  in active regions which always demonstrates very   complex topological structure as mentioned in Sect.   \ref{m_reconnection} supports the complex structure of the field as a consequence of this proposal. 
  Therefore, an observation of any  correlations between the complex thermal  pattern, higher than expected average temperature of a NS, and  the excess of the hard X ray   could (implicitly) support  the  proposed heating mechanism. 
In fact, some recent studies apparently  indicate, see review \cite{Igoshev:2021ewx} for references,  that the thermal patterns of  the NS very often   display   a complicated structure, dramatically different from  canonical poloidal dipolar magnetic field pattern.

     \subsection{Other (indirect) evidences for  DM in form of the AQN}\label{sect:paradigm}
     
  There are many hints (outside the NS physics which represents the topic of the present work) suggesting that the  annihilation events, which is inevitable feature of this framework,  may  indeed  took place in early Universe as well as   in present epoch at very different scales.  In particular, in early Universe the  AQNs do not affect BBN production for H and He, but   might be responsible for a resolution of   the  ``Primordial Lithium Puzzle" due to its   large electric charge $Z=3$, see \cite{Flambaum:2018ohm} for the details.  
  
   The very same interaction of the visible-DM components may lead to many observable effects  during the galaxy formation epoch. Indeed, while Cold Dark Matter model works very well on large scales, a number of discrepancies have arisen
between numerical simulations and observations on sub-galactic scales, see e.g. recent review 
 \cite{Tulin:2017ara} and references on original papers therein. Such discrepancies have stimulated numerous alternative proposals including, e.g. 
  Self-Interacting dark matter, Self-Annihilating dark matter,
  Decaying dark matter, and many others,
  see   \cite{Tulin:2017ara}  and references therein. Our comment here is that the AQN model represents a specific example of
  a strongly interacting    chameleon-like model: the AQNs do not interact with the surrounding material in dilute environment, but strongly interact with baryonic material in sufficiently dense environment   at the galactic scale, which helps to resolve many observed discrepancies during the  structure formation epoch   \cite{Zhitnitsky:2023znn}.

  The very same interaction of the visible-DM components may lead to large number of observable effects  also  at present epoch.   In particular, the recent studies \cite{Henry_2014,Akshaya_2018,2019MNRAS.489.1120A} suggest that there is a strong component of the diffuse far-ultraviolet (FUV) background    which is very hard  to explain by conventional physics in terms of  the dust-scattered starlight.  As argued in \cite{Zhitnitsky:2021wjb}  the  mysterious and puzzling observations of the diffuse FUV could be directly related to the annihilation events of the AQNs propagating in the galactic media.  There are numerous similar examples in many frequency bands (from radio to optical bands to UV to X rays) when the observations require an additional energy injection into the system. The AQN annihilation events 
  may provide this required (by observations) an additional source of radiation.

   We conclude this work with the following final comment.
   We advocate the idea that study of specific features of NS as mentioned in Sect.\ref{sect:tests} could shed some light on the nature of DM. It is very unexpected turn of our studies as it allows (implicitly) study the nature of the DM  by analyzing some subtle features of the NS.  
    
   The new paradigm  on the nature of cold DM (when it represented in form of macroscopical large objects as reviewed in Sect. \ref{AQN} instead of commonly accepted WIMPs) has many   consequences which are mentioned  above, and which  are consistent with all presently available cosmological, astrophysical, satellite and ground-based observations.   In fact, it may even shed some light  on the  long standing puzzles  and mysteries (outside of the NS physics) as mentioned   above and in Sect. \ref{corona}.

     \section*{Acknowledgements}
     The    motivation  for this work emerged  as a result of discussions with  Ben Safdi during  the conference
 ``Axions across boundaries between Particle Physics, Astrophysics, Cosmology and forefront Detection Technologies" which took place at the Galileo Galilei Institute  in Florence, June 2023.  I am thankful to him for the detail explanations of analysis carried out  in studies for  the hard X ray excess, which was the topic of Sect.  \ref{sect:X-ray-NS}.
 
 This research was supported in part by the Natural Sciences and Engineering
Research Council of Canada.

\appendix
\section{Magnetic Helicity ${\cal{H}}$ and its role in magnetic reconnection}\label{sect:helicity}

The main goal of this Appendix is to overview some important results on the magnetic helicity  which
is a topological invariant, and represents the observable which characterizes the dynamics of the magnetic reconnection. Needless to say that the magnetic reconnection,  according to our  proposal, plays the crucial role in transforming the static magnetic energy into the flare type events (similar to the solar flares)  which could heat the old NS.  

There are several elements which we would like to overview in this Appendix. First of all, we would like to explain 
why   magnetic helicity ${\cal{H}}$ plays a  key role in reconnection.  Secondly, we want to argue 
that  the magnetic energy related to the helicity ${\cal{H}}$ could be potentially restored   as a result of coupling  of the helicity ${\cal{H}}$ with a  reservoir  of the chirality which could  be  generated  when the  NS was sufficiently hot. 
Finally, we  also want to argue that 
 the chirality  reservoir is truly enormous in NS. 

We start with definition\footnote{In particle physics literature the magnetic helicity is defined with additional coefficient $e^2/(4\pi^2)$ in front of the integral (\ref{helicity}). This normalization factor   becomes obvious 
from eq.(\ref{total}).} of the magnetic helicity in volume $V$ which can be represented as follows \cite{choudhury}
\be
\label{helicity}
{\cal{H}}\equiv \int_V \vec{\cal{A}}\cdot \vec{{\cal{B}}}~ dV,
\ee
 where $\vec{\cal{A}}$ is the vector potential corresponding to the magnetic field $\vec{{\cal{B}}}=\vec{\nabla}\times\vec{\cal{A}}$. 
 It is known that the magnetic helicity ${\cal{H}}$ in general is not a gauge invariant observable  because the gauge potential
$\vec{\cal{A}}$ is not a gauge invariant object. However, if one  requires that the magnetic field is tangent on the surface boundary $\partial V$ of $V$, i.e. $\vec{{\cal{B}}}\cdot \vec{n}|_{\partial V}=0$, the magnetic helicity becomes well defined gauge invariant object, see e.g. \cite{choudhury}.

In simplest case when the magnetic configuration can  be represented in form of two interlinked  (but not overlapping) tubes
with fluxes $\Phi_1$ and $\Phi_2$, the magnetic helicity  ${\cal{H}}$ counts its linking number, i.e. ${\cal{H}}=2\Phi_1\Phi_2$
  is proportional to  an integer linking number if fluxes $\Phi_1$ and $\Phi_2$ are quantized. This is precisely the reason why
the magnetic helicity  is the topological invariant and cannot be easily changed during its evolution. In fact,
the crucial property of the  magnetic helicity  ${\cal{H}}$ is that it  is exactly  conserved  during the  time evolution in ideal
 MHD \cite{choudhury}. It is also known that the  magnetic helicity  ${\cal{H}}$  is odd under the $\cal{P}$   symmetry
 corresponding to : $\vec{x}\rightarrow -\vec{x}$ transformations. 
 This implies that the magnetic helicity can be only induced if there are $\cal{P}$  violating processes producing a large coherent effect on macroscopic scales.  We refer to one of the  proposals \cite{Charbonneau:2009ax} 
 with specific estimates on how it could happen. 

 
 In what follows we also need the expression for the temporal variation of magnetic helicity  as it is directly related to the dissipation rate. Differentiating of eq.  (\ref{helicity}) one arrives to
 \be
 \label{time-derivative}
 \frac{d{\cal{H}}}{dt}=-2\int_V \vec{E}\cdot \vec{{\cal{B}}} ~dV 
  \ee
  where we ignored the surface boundary term, 
  see e.g. \cite{Pariat2015},\cite{Yang2017} with explicit derivations. A key   observation  here  is    that the dissipation term in (\ref{time-derivative}) is proportional to $ \sim \vec{E}\cdot \vec{{\cal{B}}}$ which is precisely the $E\&M$ configuration  which emerges as a result of  the magnetic reconnection  as formula (\ref{E}) states. As explained in the text the induced electric field parallel to the original static magnetic field is absolutely required feature for the successful magnetic reconnection.
  The relation (\ref{time-derivative}) answers the question of why the magnetic helicity  ${\cal{H}}$ is the key player of the reconnection\footnote{Important  observation here  is that the   integrand entering  
  (\ref{time-derivative}) which  describes the dissipation of the magnetic helicity  ${\cal{H}}$   identically vanishes in ideal MHD where $\vec{E}=-\vec{v}\times \vec{{\cal{B}}}$. }.

There is one more important element here on relation between the magnetic helicity and the chirality which needs to be explained. In the chiral limit the axial current is not conserved as a result of quantum anomaly, see  e.g.   review papers \cite{Kharzeev:2009fn,Kharzeev:2015znc} on the subject, i.e.
\be
\label{chiral}
\partial_{\mu}J^{\mu}_5=\frac{e^2}{2\pi^2}\vec{E}\cdot \vec{{\cal{B}}}, 
\ee
where $J^{\mu}_5$ is the density of the axial current. In the integral form the same equation can be written as follows
\be
\label{Q_5}
\frac{dQ_5}{dt}=2\int_V\frac{e^2}{4\pi^2}\vec{E}\cdot \vec{{\cal{B}}} dV \;\; {\rm where} \;\;  Q_5\equiv \int_V J^{0}_5dV,
\ee
where the surface term has been ignored. Comparing (\ref{time-derivative}) and (\ref{Q_5}) one arrives to the following result,
see  e.g.   review papers \cite{Kharzeev:2009fn,Kharzeev:2015znc} on the subject
\be
\label{total}
\frac{d}{dt}\left(Q_5+\frac{e^2}{4\pi^2}{\cal{H}}\right)=0, 
\ee
which implies that the magnetic helicity (\ref{helicity}) in combination  with the axial charge (\ref{Q_5}) becomes  a conserved quantity, while they are not conserved separately. Important element here is that the axial current is the combination of right handed and left handed currents, i.e. $J^{\mu}_5=J^{\mu}_R-J^{\mu}_L$, while $\mu_5$ is the chemical potential for $J^{\mu}_5$. The significance of $\mu_5$ is explained in item 12 in section \ref{sect:reconnection}.  
One should comment here is that the      $\mu_5\neq 0$  is not a true chemical potential as it is not associated with any exactly conserved charges (in contrast  with  $\mu$ which  corresponds  to the conserved baryon charge).

The significance of equation (\ref{total}) is  that the $Q_5$ and ${\cal{H}}$ are strongly coupled with each other such that decrease  of $Q_5$ will lead to increase  of ${\cal{H}}$ and vice versa. This implies that the magnetic helicity ${\cal{H}}$ can be refilled and restored (in principle) after the reconnection, and the 
source of the refill  of the magnetic helicity ${\cal{H}}$ is the chiral charge $Q_5$ 
determined  by parameter $\mu_5$.  

Now we want to make several comments on the currents and their properties which could generate the chiral asymmetry and    consequently   the magnetic helicity ${\cal{H}}$, which are obviously belong to the class of the topological effects. Normally,  the  topological phenomena  are  also associated with the topological features of the sources, such as non-dissipating currents. Well known example of such relation  is the quantization of the magnetic flux and associated with the quantized  flux the non-dissipating super-current.

 It is very instructive to   explain the   differences  between the currents by analyzing their  symmetric properties. 
 We start with analysis of the conventional Ohm's law
 \be
 \label{ohm}
 \vec{J}^{\rm ohm}=\sigma \vec{E},
 \ee
where $\sigma$ is the ohmic conductivity. Both electric current $\vec{J}$ and electric field $\vec{E}$ are normal vectors ($\cal{P}$ -odd) under 
the $\cal{P}$   symmetry. Therefore, the $\sigma$ has to be $\cal{P}$ even. If we consider the time reversal symmetry 
${\cal{T}}: t\rightarrow -t$ we observe that the current  $ \vec{J}^{\rm ohm}$ is ${\cal{T}}$ odd, while the electric field $\vec{E}
=-\vec{\nabla} V$ is ${\cal{T}}$ even. Therefore, the Ohmic conductivity $\sigma$ has to be odd under the ${\cal{T}}$ reversal  for the Ohm's law (\ref{ohm}) to make sense. This is an anticipated result  since the ohmic conductivity describes processes of dissipation that produce entropy, and entropy production by the second law of thermodynamics is an irreversible process which  generates an arrow of time. In fact, all conventional  transport coefficients   are odd under  ${\cal{T}}$ reversal being consistent with presented argument. 

Now we consider the so-called Chiral Magnetic Effect (CME) when the electric current is induced due to the chiral asymmetry 
expressed in terms of the  chemical potential $\mu_5$, see reviews \cite{Kharzeev:2009fn,Kharzeev:2015znc}:
 \be
 \label{top}
 \vec{J}^{\rm top}=\sigma_5 \vec{\cal{B}}, \;\;\;\;\;  \sigma_5=\frac{e^2}{2\pi^2}\mu_5.
 \ee
 The difference with previous case of the Ohm's law (\ref{ohm}) is that  the magnetic field $\vec{{\cal{B}}}=\vec{\nabla}\times \vec{\cal{A}}$ is ${\cal{T}}$ odd
 because the vector potential $ \vec{\cal{A}}$ is ${\cal{T}}$ odd. From (\ref{top}) we infer that $\sigma_5$ has to be  ${\cal{T}}$ even, and the topological current  $\vec{J}^{\rm top}$ entering (\ref{top}) is expected to be non-dissipating. 
 
 It is very instructive to compare this analysis with another type of  non-dissipating current which is  induced in superconducting materials. This 
 is also important from phenomenological viewpoint  as the NS is  believed to be a large superconductor\footnote{It remains to be a matter of debates whether superconductivity realized in NS is the type -I or type II  superconductor \cite{Buckley:2003zf}.}.
 The corresponding physics is captured by the London relation between
the electric current and gauge potential
  \beq
 \label{London}
 \vec{J}^{\rm London}= \lambda^{-2} \vec{\cal{A}},    \;\;\;\;\;   \vec{\nabla}\cdot \vec{\cal{A}}=0, 
 \eeq
 where $\lambda$ is the penetration length. 
 The vector potential $ \vec{\cal{A}}$ as well as $\vec{J}^{\rm London}$ are  ${\cal{T}}$ odd functions, which suggests that the London current $\vec{J}^{\rm London}$ is non-dissipating. Indeed, the magnetic flux through
an Abrikosov vortex for type II superconductor  is quantized. This means
that the circulating super-current is topologically protected-- it is not allowed to dissipate as the flux is quantized. 
 
 \exclude{
 Indeed, in the absence of the electric charges  
 $\vec{E}=- \partial/\partial t(\vec{\cal{{A}}})$ and the London equation (\ref{London} can be written as 
  \be
 \label{London1}
\vec{E} = \lambda^2\dot{\vec{J}}^{\rm London},   
 \ee
 which obviously implies that  a constant electric field induces an electric current growing with time. 
 }
 The non-dissipating nature of the topological current (\ref{top}) is also supported by analysis 
   \cite{Charbonneau:2009ax} where it has been argued that the conventional ${\cal{P}}-$ and ${\cal{T}}$-even QED processes (which are normally incorporated in MHD analysis) cannot eliminate the current (\ref{top}). Indeed,  the correlation 
   $\la  \vec{J}^{\rm top} \cdot \vec{\cal{B}} \ra$ is ${\cal{P}}$ -odd  correlation which cannot be changed by conventional ${\cal{P}}$ -even QED processes\footnote{We emphasize that the claim is not that the transitions $L\rightarrow R$ and $R\rightarrow L$ do not occur all the time. These transitions of course occur in QED as for example, the mass term flips the chirality. The claim is that the expectation value $\la  \vec{J}^{\rm top} \cdot \vec{\cal{B}} \ra$  in equilibrium (including magnetic  portion of the helicity $\cal{H}$) cannot be washed out without ${\cal{P}}$ -odd weak interactions.}. 
   The ${\cal{P}}$ odd weak interactions are capable to  diminish the induced current (\ref{top})  which was generated by weak interactions when the NS's temperature was sufficiently high, see  \cite{Charbonneau:2009ax} for the details. The surface effects can also play a role as helicity may leak through the surface. However, these surface effects are expected to be subdominant, and not considered here. 
   
   We finish this Appendix with order of magnitude estimate of the total energy reservoir which potentially can refill the magnetic helicity.  As we mentioned after (\ref{total}) the strength of helicity ${\cal{H}}$ can be restored after every event of reconnection by conversion of the energy from $Q_5$. By definition, $J^{0}_5=\int_V(J^{0}_R-J^{0}_L) dV$ is the difference between right handed and left handed densities. The source of this asymmetry is $\mu_5$, which was estimated in  \cite{Charbonneau:2009ax}. The most important parameters for the present estimate  from  \cite{Charbonneau:2009ax} is the asymmetry parameter $P_{\rm asym}$ and the Fermi momenta $k_e\sim \mu$ for electrons   numerically assume the following values:
   \be
   \label{asym}
   P_{\rm asym}\sim \frac{\mu_5}{\mu}\approx 2\cdot 10^{-5}, \;\;\;\;\;\ \mu\approx 100 \rm ~MeV.
   \ee
   From (\ref{asym}) one can estimate the total difference between right handed and left handed electrons in the entire NS as follows:
   \beq
   \label{Q5}
  \left(\frac{\mu_R^3}{3\pi^2} -  \frac{\mu_L^3}{3\pi^2} \right)\left( \frac{4\pi R^3}{3}\right)\approx\frac{\mu^2\mu_5}{\pi^2}\left( \frac{4\pi R^3}{3}\right)\sim 10^{51}.~~~~
   \eeq
   The total energy reservoir which is available to refill the magnetic helicity ${\cal{H}}$  can be estimated by multiplication eq. (\ref{Q5}) to a typical  energy of electrons, which is determined by $\mu$. Therefore, we arrive  to the following estimate for the total available energy 
   \beq
   \label{reservoir}  
   E^{\rm tot}_{(R-L)}\sim \mu\cdot \left(\frac{\mu^2\mu_5}{\pi^2}\right)\cdot\left( \frac{4\pi R^3}{3}\right)\sim 10^{47} \rm erg.
   \eeq
  
  It is instructive to compare this total  energy (produced  at early times due to the generation of the right-left asymmetry) with total magnetic energy as given by (\ref{m_energy}). It is clear that the total energy reservoir (\ref{reservoir}) is enormous. This amount of energy is more than sufficient to refill    the magnetic helicity ${\cal{H}}$ after each magnetic reconnection to heat the NS surface, which is precisely the proposal
  mentioned  in items  11-14 in Sect. \ref{sect:reconnection}.  As we mentioned  above this asymmetry cannot be washed out by conventional QED $\cal{P}$ even processes incorporated into MHD. Only  the $\cal{P}$ odd weak interactions are capable to diminish this asymmetry. But these processes are very slow at low temperatures \cite{Charbonneau:2009ax}, and unlikely to play a role.  

  \section{\label{radio} Radio bands as indicator of the magnetic reconnection}
  
                     It is known that the magnetic reconnection is also accompanied by emission in radio frequency bands as the solar flare observations suggest.  Therefore, one could expect that the search  in radio frequency  bands could be an alternative way to study the heating mechanism due to the magnetic reconnection in NS along with X ray as suggested in Sect. \ref{x-ray}. 
      The main goal of this Appendix is to make an order of magnitude estimates for expected the radio signal from NS which accompanies  the magnetic reconnection. We follow the same logic in these estimates as we used previously for the X ray analysis in Sect. \ref{x-ray}. 
  
   It is known that along with X rays there is  another indicator for solar flares, the so-called $F_{10.7}$ flux which reflects the solar activity in radio wave band with wave length $10.7$cm. Typical enhancement during the solar flare is about 200 sfu (solar flux units),
        see Fig 2 in   ref.\cite{Bruevich_2017}. Converting  sfu to conventional units we arrive to the following estimate for the radio emission during the solar flare
          \beq
            \label{ratio2}
           \la F^{\odot}_{\rm radio}\ra&\approx& 6\cdot 10^{-8}  {\rm \frac{erg}{s\cdot cm^2}}, \;\;\;\; \rm sfu\equiv 10^{-22}\rm \frac{W}{m^2 Hz} \nonumber \\
             \xi^{2.7 \rm GHz}_{\rm radio} &\equiv&\frac{ \la F^{\odot}_{\rm radio}\ra}{\la F^{\odot}_{X}\ra }\approx 6\cdot 10^{-6}.
            \eeq
            We assume that this ratio $ \xi_{\rm radio}$ holds for NS as well. Therefore, assuming the flat spectrum  we arrive to the following estimate
            for $\la F^{\rm NS}_{\rm radio}\ra$:
  \beq
\label{radio-NS}
&&\la F^{\rm NS}_{\rm radio}\ra\sim  \xi^{1.4 \rm GHz}_{\rm radio} \cdot \la F^{\rm NS}_{\rm X}\ra  \\
&\sim& 3\cdot  (10^{-19}-10^{-20}) \left(\frac{T_s}{10^6 \rm K}\right)^4 \left(\frac{0.1 \rm  kpc}{r}\right)^2{\rm \frac{erg}{s\cdot cm^2}},\nonumber
\eeq
 where we  use our estimate for $\la F^{\rm NS}_{\rm X}\ra$ from (\ref{X_ray}).
 
  It is instructive to present a numerical value of  the flux $\la F^{\rm J1856}_{\rm radio}\ra$ for $J1856$  from M7 stars where X ray flux is  recorded  according to eq. (\ref{observations}). 
    \beq
     \label{radio-prediction}
 \la F^{\rm J1856}_{\rm radio,~1.4 \rm GHz}\ra \sim 4\cdot 10^{-21}  \rm \frac{erg}{s\cdot cm^2}, \;\;\; [\rm prediction], \;\;\;
 \eeq   
  while the observed upper limit for this star in $\rm 1.4~ GHz$ band is $10^{-4} \rm (mJy) (kpc^2)$, which is almost two orders of magnitude higher than (\ref{radio-prediction}), see Fig. 14 in  \cite{Dessert:2019dos}.   Therefore, we are not optimistic with possible observations of the radio signals from NS which always accompany the   magnetic reconnection, but we quite optimistic with possible observations in hard X ray energy band as discussed in the main body of the text in Sect. \ref{x-ray}.

 \exclude{
\appendix
  \section{\label{SP}Sweet-Parker's (SP) theory }

 We start by introducing the most important parameters of the problem
 \be
 \label{definitions}
 S=\frac{Lv_A}{\chi_m}, ~~~~~ v_A=\frac{\cal{B}}{\sqrt{4\pi\rho}}, ~~~~~ \chi_m=\frac{c^2}{4\pi\sigma}, 
 \ee
where  $S$ is the so-called the Lundquist number,   $L$ is the typical size of the problem,   $v_A$ is Alfv\'{e}n speed, $\rho$ is the plasma's mass density,  $\chi_m$ is the magnetic diffusivity, and finally  $\sigma$ is the electrical conductivity of the plasma. 
The most important parameter for our future estimates is the dimensionless parameter $S$ which assumes the following values for typical coronal conditions, $S\sim (10^{12}-10^{14})$. 

Original idea on magnetic reconnection was formulated   by Sweet \cite{Sweet}  and Parker \cite{Parker1} sixty years ago.
Using simple dimensional arguments, Sweet and Parker (SP) have shown that the reconnection time $\tau_{\rm rec}$ is  
quite slow and expressed in terms of the original parameters of the system as follows
\beq
\label{eq:SP}
\frac{\tau_A}{\tau_{\rm rec}}\sim \frac{1}{\sqrt{S}}, ~~~~ \tau_A\equiv\frac{L}{v_A}, ~~~~ \frac{u_{\rm in}}{u_{\rm out}}\sim  \frac{1}{\sqrt{S}},  \nonumber \\
~~~~ \frac{l}{L}\sim \frac{1}{\sqrt{S}}, ~~~~~\tau_{\rm rec}\sim \frac{L}{u_{\rm in }}
\eeq
 where $u_{\rm in}$ is the  velocity of reconnection between oppositely directed fluxes of thickness $l$, and  $u_{\rm out}\sim v_A$ is normally assumed to be of   order of  the Alfv\'{e}n velocity. The scaling relations (\ref{eq:SP}) predicted by SP theory are  obviously insufficient to explain the reconnection rates observed in corona due to the very large numerical values of $S\sim (10^{12}-10^{14})$.
 
 The next step to speed up the reconnection rate has been undertaken in \cite{Petschek} with some important amendments in \cite{Kulsrud} where it was  argued that the reconnection rate could be much faster than the  original formula (\ref{eq:SP}) suggests.
 However, some subtleties remained in the proposal \cite{Petschek,Kulsrud}. Furthermore, the numerical simulations reproduce conventional scaling formula    (\ref{eq:SP}) at least for moderately large $S\lesssim 10^4$. In the last 10-15 years 
  large number of new ideas have been pushed forward. It includes, but not limited to such processes as plasmoid- induced reconnection,  fractal reconnection, to name just a  few. 
 
 It is not the goal of the present work to analyze  the assumptions, justifications, and the problems related to proposals \cite{Sweet,Parker1,Petschek,Kulsrud},  and we refer to
 the recent review papers \cite{Shibata:2016,Loureiro} for recent developments and relevant discussions  on these matters. We are only underlying that AQNs acting as triggers (through shock waves) can drastically speed up the reconnection as     reviewed in previous section, and provide a physical context consistent with the ideas of \cite{kumar_1,kumar_2}. Similarly,  as shown in \cite{Tanuma}, the shock waves may trigger and ignite (in very different circumstances) sufficiently fast reconnections\footnote{\label{MHD}Furthermore, the 2d MHD simulations \cite{Tanuma} show  that  a large number of different phenomena,
    including SP reconnection \cite{Sweet, Parker1}, Petschek reconnection \cite{Petschek, Kulsrud}, tearing instability, formation of the magnetic islands, and many others, may all take place at different phases in  the evolution of the system, see also reviews \cite{Shibata:2016,Loureiro}.}. 

There are few  important parameters which control  the dynamics of the  system: in addition  to $S$  it is convenient to introduce another dimensionless  parameter $\beta$ defined by (\ref{beta}) which determines the importance of the magnetic pressure in comparison with the gas pressure,

The idea   that the shock waves may drastically increase the rate of magnetic reconnection is not new, and has been discussed previously in the literature \cite{Tanuma}, though a quite different context: it was applied to interstellar medium in the presence of a supernova shock. \cite{Tanuma} showed that the shock waves may trigger and ignite     sufficiently fast reconnections.
 }

   
  \bibliography{NS-cooling-1}

\end{document}